\documentclass[twocolumn, appendixfloats, revtex4, numberedappendix, twocolappendix, iop]{openjournal}

\usepackage{amsmath, amsfonts, amssymb} % Additional maths symbols
\usepackage{graphicx} % Allows the use of adding images, or something like that
\usepackage[colorlinks, linkcolor=blue, citecolor=blue]{hyperref} % This sometimes goes mad if an equation is broken over a page and pdf latex exits with some unintelligible error
\usepackage{xspace} % \xspace which is useful for intelligent spacing after newly defined commands
\usepackage{xcolor} % Enables coloured text
\usepackage{upgreek} % For upright greek characters
\usepackage{tabularx} % ?

\usepackage{orcidlink}

\newcommand{\ba}{\begin{eqnarray}}
\newcommand{\ea}{\end{eqnarray}}
\newcommand{\be}{\begin{equation}}  
\newcommand{\ee}{\end{equation}}

\newcommand{\dn}{\ifmmode D_{\rm n}4000 \else $D_{\rm n}4000$\fi\xspace}
\newcommand{\Ha}{\ifmmode \mathrm{H}\,\alpha \else H\,$\alpha$\fi\xspace}
\newcommand{\Hb}{\ifmmode \mathrm{H}\,\beta \else H\,$\beta$\fi\xspace}
\newcommand{\Hd}{\ifmmode \mathrm{H}\,\delta \else H\,$\delta$\fi\xspace}
\newcommand{\Hda}{\ifmmode \mathrm{H}\,\delta_{\rm A} \else H\,$\delta_{\rm A}$\fi\xspace}
\newcommand{\Hdf}{\ifmmode \mathrm{H}\,\delta_{\rm F} \else H\,$\delta_{\rm F}$\fi\xspace}
\newcommand{\LHa}{\ifmmode L_{\Ha} \else $L_{\Ha}$\fi\xspace}
\newcommand{\LFIR}{\ifmmode L_\mathrm{FIR} \else $L_\mathrm{FIR}$\fi\xspace}
\newcommand{\LTIR}{\ifmmode L_\mathrm{TIR} \else $L_\mathrm{TIR}$\fi\xspace}
\newcommand{\Ldust}{\ifmmode L_\mathrm{TIR} \else $L_\mathrm{TIR}$\fi\xspace}
\newcommand{\HII}{\ifmmode \mathrm{H}\,\textsc{ii} \else H~{\sc ii}\fi\xspace}
\newcommand{\Mgal}{\ifmmode M_* \else $M_*$\fi\xspace}
\newcommand{\nii}{\ifmmode [\mathrm{N}\,\textsc{ii}] \else [N~{\sc ii}]\fi\xspace}
\newcommand{\oii}{\ifmmode [\mathrm{O}\,\textsc{ii}] \else [O~{\sc ii}]\fi\xspace}
\newcommand{\oiii}{\ifmmode [\mathrm{O}\,\textsc{iii}] \else [O~{\sc iii}]\fi\xspace}
\newcommand{\WHa}{\ifmmode W_{\Ha}\else $W_{\Ha}$\fi\xspace}
\newcommand{\eg}{e.g.\@\xspace}
\newcommand{\ie}{i.e.\@\xspace}

\newcommand{\starlight}{\textsc{starlight}\xspace}

\newcommand{\magphys}{\textsc{magphys}\xspace}
\def\mnras{MNRAS}
\def\apj{ApJ}
\def\apjl{ApJ Let.}

\def\aap{A\&A}
\def\apjs{ApJS}

\begin{document} % OJA

%\title{How much does optical spectroscopy miss? A multi-wavelength study of galaxies with strong Balmer absorption lines} % OJA
\title{The infrared luminosity of retired and post-starburst galaxies: A cautionary tale for star formation rate measurements}

% Authors 
% For ArXiv submission  \vspace{-1.3cm} needs to go IN the author list
\author{\vspace{-1.3cm} Vivienne Wild\,\orcidlink{0000-0002-8956-7024}$^{1}$\footnote{vw8@st-andrews.ac.uk},
Natalia Vale Asari\,\orcidlink{0000-0003-0842-8688}$^{2}$, 
Kate Rowlands\,\orcidlink{0000-0001-7883-8434}$^{3}$, 
Sara L. Ellison\,\orcidlink{0000-0002-1768-1899}$^{4}$,\\
Ho-Hin Leung\,\orcidlink{0000-0003-0486-5178}$^{1}$,
Christy Tremonti\,\orcidlink{0000-0003-3097-5178}$^{5}$
}
\affiliation{
% List of institutions
$^{1}$School of Physics \& Astronomy, University of St Andrews, North Haugh, St Andrews, KY16 9SS, U.K.\\
$^{2}$Departamento de F\'{\i}sica--CFM, Universidade Federal de Santa Catarina, C.P.\ 5064, 88035-972, Florian\'opolis, SC, Brazil\\
$^{3}$AURA for ESA, Space Telescope Science Institute, 3700 San Martin Drive, Baltimore, MD 21218, USA; William H. Miller III Department of Physics and Astronomy, Johns Hopkins University, Baltimore, MD 21218, USA\\
$^{4}$Department of Physics \& Astronomy, University of Victoria, Finnerty Road, Victoria, British Columbia, V8P 1A1, Canada\\
$^{5}$Department of Astronomy, University of Wisconsin-Madison, Madison, WI 53706, USA
}

% Abstract
\begin{abstract}
In galaxies with significant ongoing star formation there is an
impressively tight correlation between total infrared luminosity
(\LTIR) and \Ha luminosity (\LHa), when \Ha is properly corrected for
stellar absorption and dust attenuation. This long-standing result
gives confidence that both measurements provide accurate estimates of
a galaxy's star formation rate (SFR), despite their differing
origins. To test the extent to which this holds in galaxies with lower
specific SFR ($\mathrm{sSFR} = \mathrm{SFR}/\Mgal$, where \Mgal is the
stellar mass), we combine optical spectroscopy from the Sloan Digital
Sky Survey (SDSS) with multi-wavelength (FUV to FIR) photometric
observations from the Galaxy And Mass Assembly survey (GAMA). We find
that \LTIR/\LHa increases steadily with decreasing \Ha equivalent
width (\WHa, a proxy for sSFR), indicating that both luminosities
cannot provide a valid measurement of SFR in galaxies below the
canonical star-forming  sequence. For both `retired galaxies' and
`post-starburst galaxies', \LTIR/\LHa can be up to a factor of 30
larger than for star-forming galaxies. The smooth change in
\LTIR/\LHa, irrespective of star formation history, ionisation or
heating source, dust temperature or other properties, suggests that
the value of \LTIR/\LHa is determined by the balance between star-forming
regions and ambient interstellar medium contributing to both \LTIR and
\LHa. It is not a result of the differing timescales of star formation that these luminosities probe. While \LHa can only be used to estimate the SFR for galaxies
with $\WHa > 3$~\AA\ ($\mathrm{sSFR} \gtrsim 10^{-11.5}$/yr), we argue
that the mid- and far-infrared can only be used to estimate the SFR of
galaxies on the star-forming sequence, and in particular only for
galaxies with $\WHa >10$\,\AA\ ($\mathrm{sSFR} \gtrsim
10^{-10.5}$/yr). We find no evidence for dust obscured star-formation
in local post-starburst galaxies.

\end{abstract}


\maketitle % OJA
%\tableofcontents

\section{Introduction}
Measuring the rate at which galaxies form stars is one of the key tools used to understand how they form and evolve. The field has benefited from a wide range of independent methods, from data spanning the full electro-magnetic spectrum, allowing for internal consistency checks and validation \citep[see][for a review]{KennicuttEvans2012}. This has led to significant confidence amongst extra-galactic astronomers in our ability to accurately measure the star formation rate (SFR) of any galaxy, of any type, at any redshift, with important caveats often under-appreciated. 

Here we focus on one particular relation, between total infrared (TIR) luminosity (including mid- and far-IR) and \Ha luminosity. Once \Ha has been properly corrected for stellar photospheric Balmer absorption and for dust attenuation via the Balmer decrement method, the two luminosities are impressively tightly correlated in local star-forming galaxies \citep{RosaGonzalez2002, kewley2002b, Rosario2016}, despite their different physical origins.  Balmer emission lines are produced via recombination of electrons to the $n=2$ level of the hydrogen atom; in star-forming galaxies this predominantly occurs following photoionisation and subsequent recombination and cascade within the \HII regions that surround hot (O and B type) stars. Mid to far-IR emission is caused by the heating of dust grains through the absorption of ultra-violet (UV) to optical photons. In highly star-forming galaxies these photons come predominantly from young, hot stars and interact with the dense, dust rich birth clouds of the star forming regions, leading to the strong relation between TIR emission and SFR. Unfortunately, neither of these SFR estimators are free of contamination or complications. 

The basic physics of Balmer emission in star-forming galaxies is simple and well understood, with only photons with energies greater than 13.6\,eV (Lyman continuum photons) able to contribute to the ionisation of the hydrogen atoms in the first place, meaning O and B stars are the primary contributor to Balmer emission in star-forming galaxies. However, there are two key complications in converting a \Ha emission line luminosity into a SFR. These are the attenuation of the photons as they travel through the galaxy ISM once leaving the \HII\ regions, and the contribution to the far-UV interstellar radiation field (ISRF) from active galactic nuclei (AGN), shocks and hot low-mass evolved stars (HOLMES; \citealp{FloresFajardo.etal.2011a}), such as hot post-asymptotic giant branch stars and white dwarfs \citep{stasinska2008, CidFernandes2011}. Through modelling the stellar continuum in SDSS galaxies using population synthesis models, \citet{CidFernandes2011} showed that galaxies with a low equivalent width of the \Ha\ emission line ($\WHa < 3$\,\AA) had an ionisation source consistent with the HOLMES present in the galaxy continuum spectrum, calling them `retired galaxies'. In these galaxies \LHa\ is not proportional to their SFR, but is proportional to the total stellar mass of lower mass stars available to provide the ionising photons, as well as the presence of a reservoir of warm gas \citep[see][for a comparison of retired galaxies with and without \Ha emission]{herpich2018}.

Once the \Ha photon is emitted from the \HII region close to the site of star formation, it can be absorbed and scattered by dust within the galaxy before being received by our telescopes\footnote{We are ignoring here the possibility that dust grains within the \HII region might absorb the photon before it ever reaches the hydrogen atoms to ionise them. Such competition for Lyman continuum photons is difficult to detect \citep[\eg][]{CharlotLonghetti2001, Murphy2011}.}. In the case that \emph{some} of the line emission still escapes and the dust distribution in the galaxy is smooth, the loss of light can be corrected for via the known ratio between Balmer line fluxes, combined with a knowledge of the dust attenuation as a function of wavelength \citep[the Balmer decrement method, see \eg][]{Groves2012}. In the case of clumpy dust the detected Balmer emission is weighted towards lower dust regions, and therefore the dust attenuation in the galaxy is underestimated using this method \citep{ValeAsari2020}.  Comparison between spatially resolved and integrated light observations indicate that this effect is at the $\lesssim10$\% level in the majority of galaxies \citep{ValeAsari2020, Belfiore2023}, however the relative lack of very high spatial resolution integral field spectroscopy makes this estimate uncertain. At the extreme end of this scenario, there is clearly no way to correct for Balmer emission that has been entirely blocked by dense dust clouds. 

On the other hand, the physics of mid- to far-IR luminosity is highly complex, with dependence on the gas density, dust-star geometry, the distribution of dust grain sizes and types which can vary with metallicity, and the shape of the ISRF, which depends on the stellar population balance \citep[\eg][]{DraineLi2007, Paladini2007, Bernard2010, Nersesian2019}. It is predominantly the strong correlation with other measurements that provides the evidence for the link between TIR and SFR in high SFR galaxies. While Balmer emission lines are somewhat impacted  by HOLMES, significant dust heating is caused by all stars with masses below O and B type stars. Diffuse dust emission permeates the entire galaxy contributing significantly to the IR luminosity of galaxies or regions with lower ongoing SFRs \citep[\eg][]{Helou1986, LonsdalePersson1987, Paladini2007, Bendo2010, Kennicutt2011, Calzetti2013book, Nersesian2019}. As a result, the TIR to \Ha luminosity ratio (\Ldust/\LHa) has been observed to increase with decreasing specific SFR ($\mathrm{sSFR}=\mathrm{SFR}/\Mgal$, where \Mgal is a galaxy's stellar mass) in integrated SDSS--Herschel \citep{Rosario2016} and SDSS--WISE \citep{Salim2016} comparisons, and spatially resolved PHANGS--MUSE comparisons \citep{Belfiore2023}. While \citet{Rosario2016} put the effect down to selection bias, \citet{Belfiore2023} interpreted it as due to contamination of the IR luminosity by diffuse ISM dust.

The fact that there may be significant reservoirs of dust invisible to optical astronomers due to the enormous optical depths of stellar birth clouds has led to the idea that FIR may in fact be a \emph{better} indicator of SFR in certain types of galaxies. This has become a critical discussion in particular in relation to galaxies which are currently shutting down their star formation (so-called `quenching'), and has crucial importance as the field grapples with understanding the wide range of processes that cause galaxies to stop forming stars. If the dust in a galaxy is sufficiently dense to bury light from a significant number of the O and B stars, as well as their surrounding \HII regions, then we have the case of a `dust obscured starburst'. Such a galaxy may reveal its presence in the optical via strong Balmer absorption lines, as the longer living A and F stars emerge from the dense clouds shrouding the younger O and B stars \citep{smail1999,PoggiantiWu2000,Geach2009,Baron2022}. 

However, an excess Balmer absorption line strength is exactly the signature used to detect galaxies which have recently and rapidly shut down their star formation. Ironically, it is in these rapidly quenching galaxies where we might also expect to see the largest fraction of IR dust emission coming from the ambient ISM, unrelated to current star formation, as clearly demonstrated using hydrodynamical merger simulations by \citet{Hayward2014}. This is because (a) we expect the diffuse dust emission located far from previous sites of star formation to still be present, (b) there is significant UV--optical light emitted by the A and F stars, (c) these A and F stars may also still be co-located with the remnant dust from the starburst episode. So while \Ha disappears rapidly following the starburst, total IR luminosity is expected to remain high, leading to an order of magnitude over-estimation in star formation, if using IR luminosity, from the point at which star formation starts to decline, right into the quenched phase many 100's of Myr later (see figures~2 and 3 of \citealt{Hayward2014}). 

As the debate about the role of total IR emission as a SFR indicator in quenching galaxies continues, and the origin of IR emission in `retired' galaxies has not yet been studied, this paper takes a fresh look at the relationship between total IR luminosity and \Ha luminosity in a wide range of galaxies, including star-forming, post-starburst and retired galaxies. In Section~\ref{sec:sample} we present the catalogues that we have combined in our analysis, in Section~\ref{sec:methods} we summarise the relevant methods, and in Section~\ref{sec:results} we present our results. We discuss our results with the help of a simple toy model in Section~\ref{sec:discussion} and summarise in Section~\ref{sec:summary}. Throughout the paper we assume a cosmology of $\Omega_M=0.3$, $\Omega_\Lambda=0.7$ and $h=0.7$ when converting fluxes into luminosity. All derived catalogues that we use assume a \citet{Chabrier2003} initial mass function for conversion of stellar light into stellar masses. Unless otherwise stated \LHa refers to dust attenuation and aperture corrected \Ha luminosity.

\section{Data}\label{sec:sample}
\begin{figure}
    \centering
    \includegraphics[width=\linewidth, trim=10 10 0 0]{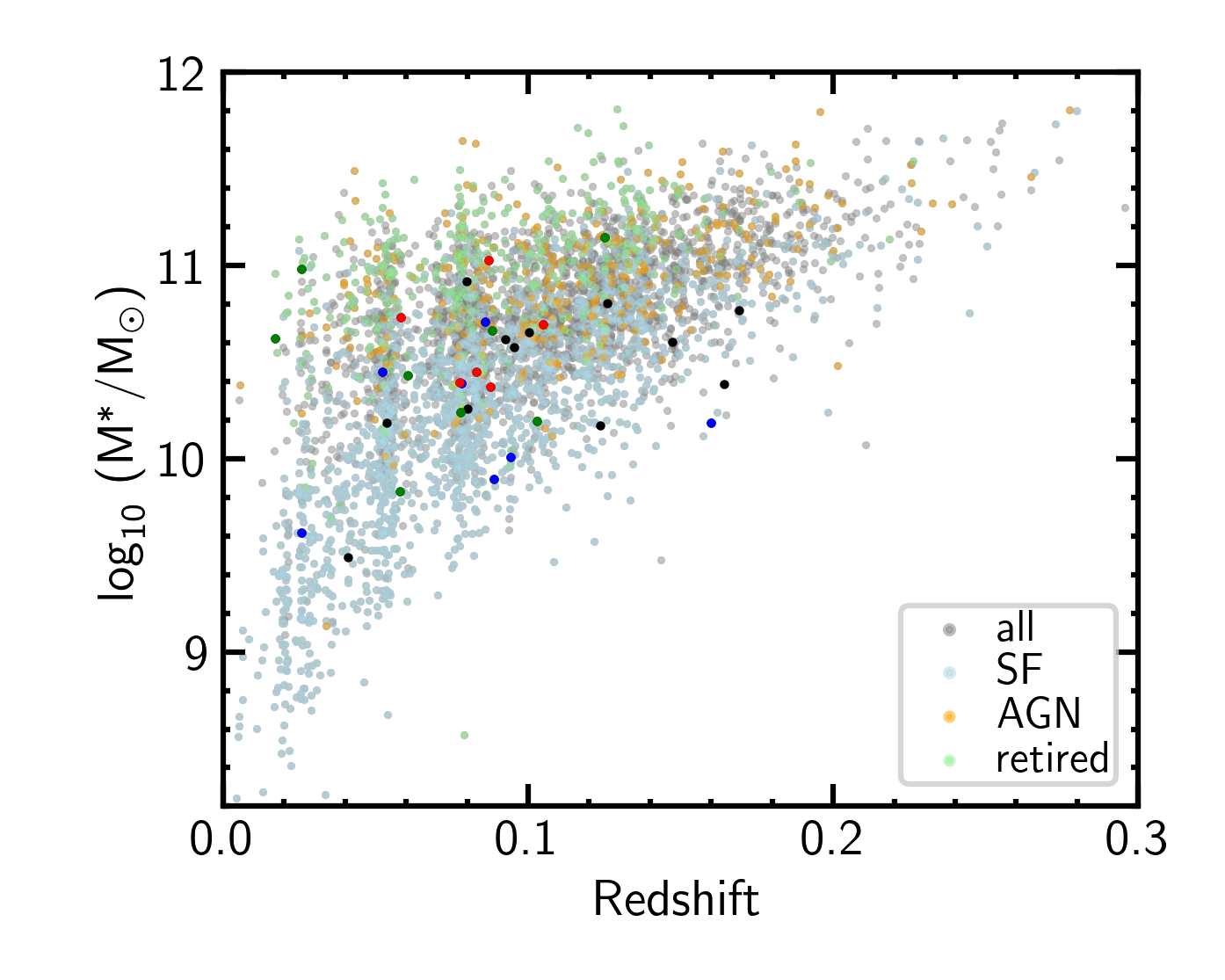}
    \includegraphics[width=\linewidth, trim=10 10 0 0]{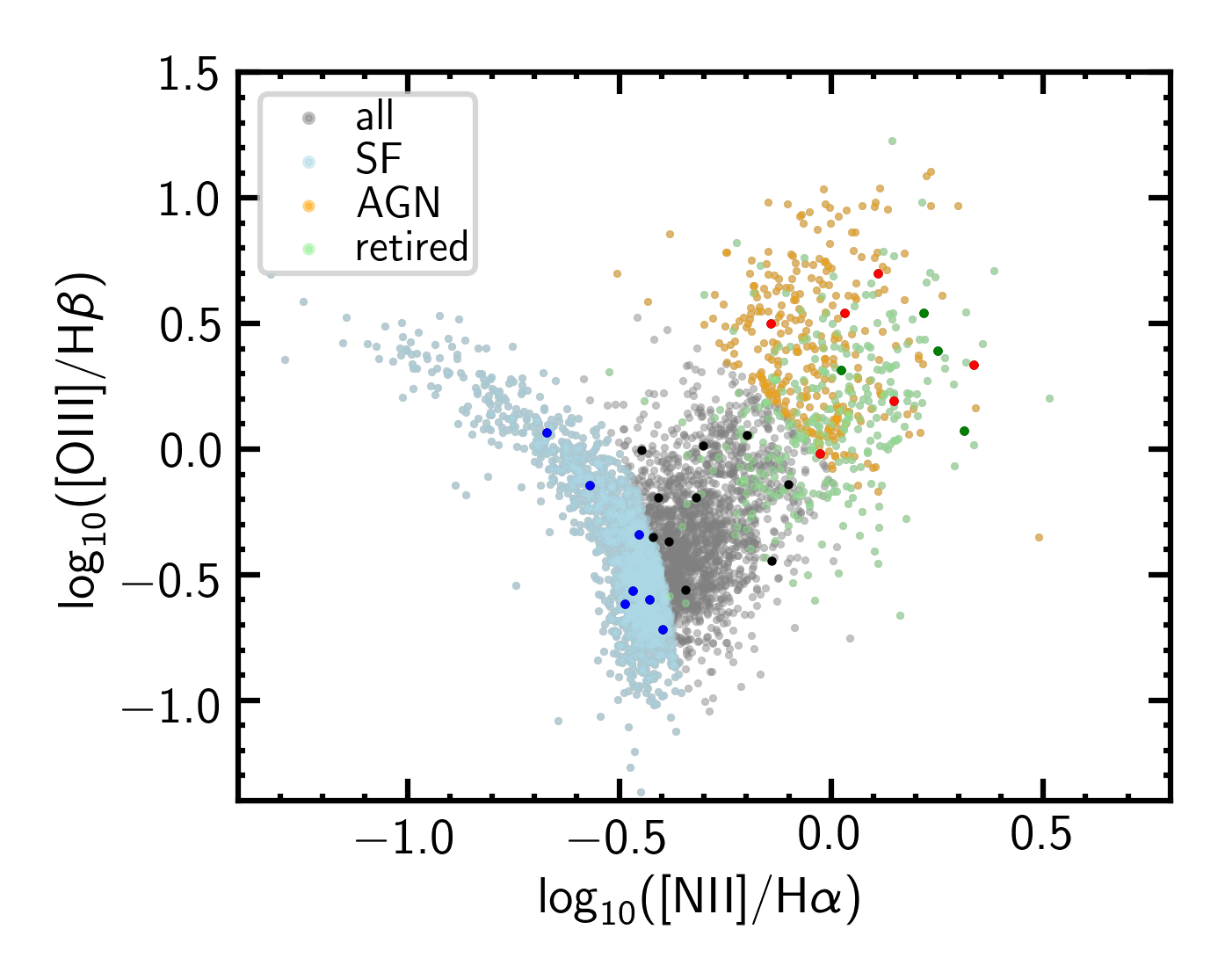}
    \includegraphics[width=\linewidth, trim=10 10 0 0]{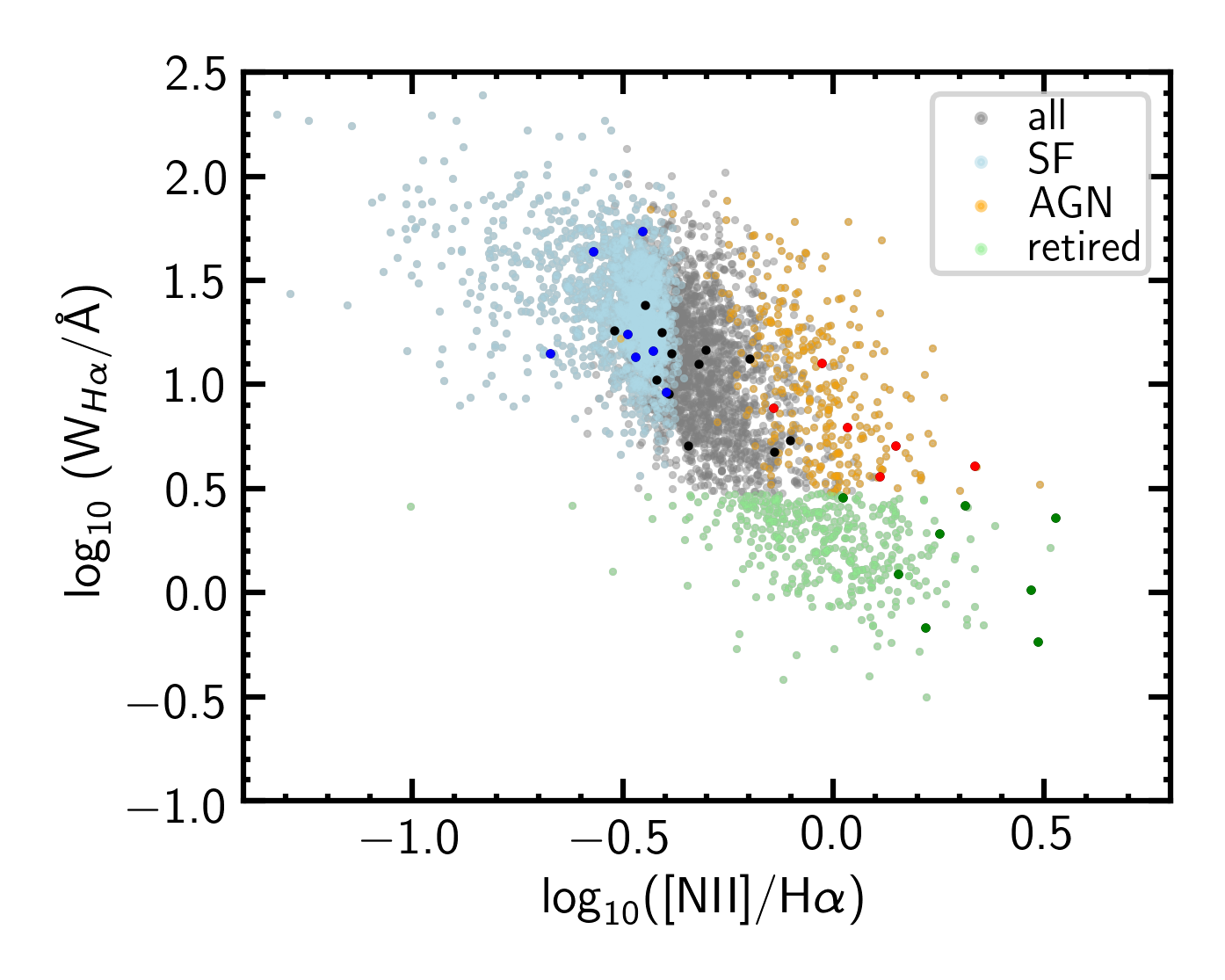}
    \caption{A summary of the properties of the galaxies studied in this paper. Galaxies are colour coded according to their emission line properties into star-forming (SF), AGN and retired galaxies, as described in the text. Post-starburst galaxies are identified via their stellar continuum properties, and are shown as darker coloured circles.  {\it Top:} redshift vs.\ total stellar mass. {\it Centre:} \nii/\Ha vs.\  \oiii/\Hb line flux ratios. {\it Bottom:} \nii/\Ha line flux ratio vs.\ the equivalent width of the \Ha emission line.  
    }
    \label{fig:sample}
\end{figure}

Our analysis requires a large sample of galaxies with good quality optical continuum spectroscopy to provide continuum subtracted, dust attenuation corrected \Ha luminosity, combined with aperture-matched UV, optical and mid- to far-IR multiwavelength photometric observations to provide accurate dust luminosities. To this end, we combine the spectroscopic galaxy sample from the Sloan Digital Sky Survey (SDSS) Data Release 7 \citep[DR7,][]{SDSSDR72009}, with photometry from the Galaxy And Mass Assembly (GAMA) Data Release 4 equatorial survey \citep{Driver2022}. 

From SDSS we take nebular emission line fluxes, errors and equivalent widths from the \starlight--SDSS catalogue (\citealp{CidFernandes.etal.2009a}; see also \citealp{Mateus.etal.2006a, Stasinska2006, Asari.etal.2007a, CidFernandes.etal.2010a}). The underlying stellar continuum is subtracted prior to measurement of the emission line fluxes using the \starlight full spectral fitting package \citep{CidFernandes.etal.2005a}.  Stellar masses are calculated from the mass-to-light ratio of the \starlight model, and corrected for aperture loss using the fiber-to-total $z$-band luminosity fraction. The luminosity of the \Ha emission line is corrected for dust attenuation using the \citet{Cardelli.Clayton.Mathis.1989a} law with $R_V = 3.1$ and assuming an intrinsic Balmer decrement of $\Ha/\Hb = 2.86$. This steeper dust law than \eg \citet{Calzetti2000} is appropriate for correcting emission lines, due to the screen-like geometry of the dust in \HII regions \citep{wild2007, daCunha2008_magphys, wild2011}. While the assumed dust law alters the absolute conversion between \LHa and \Ldust in the case that both trace SFR, this does not impact the conclusions of this paper. 

From GAMA we obtain 21 band multi-wavelength photometry from the {\sc lambdar} catalogue \citep{LambdaR_2016}, which is aperture-matched and deblended, with variations in the point-spread function and pixel scale across the various bands properly accounted for \citep[see also][]{Driver2016_GAMA}. Most importantly for this paper, the GAMA equatorial fields include observations from Data Release 1 of the Herschel--ATLAS far-IR space telescope survey \citep[H-ATLAS,][]{HAtlasDR1_2016a, HAtlasDR1_2016b}, and the Wide Field Infrared Survey Explorer which covers the mid-IR \citep[WISE, ][]{WISE2010, Cluver2014_WISE}. These two surveys provide 6 band photometry across the crucial 22--500\,\textmu{}m wavelength range\footnote{\ie\ Wise W4, PACS 100, PACS 160, SPIRE 250, SPIRE 350, SPIRE 500.} required to measure and characterise the emission from dust in galaxies. The majority of the galaxies in our sample also have GALEX NUV and FUV fluxes detected at $>3\sigma$, including the post-starburst and retired galaxies. The latter are typically detected with lower significance. We make use of the GAMA DR4 \magphys\ catalogue, which takes input photometry from the {\sc lambdar} catalogue using the code of \citet{daCunha2008_magphys} to measure stellar population and dust properties for GAMA galaxies.

\subsection{Sample selection}
From the SDSS spectroscopic main galaxy sample \citep[{\sc primtarget}\,$=64$,][]{strauss2002} we exclude a small number of objects for which the SDSS 1D spectroscopic pipeline has identified problems, \ie\ if the {\sc z\_warning} flag is set, or the velocity dispersion is fitted to the maximum value of {\sc v\_disp}$=850$\,kms$^{-1}$ \citep{SDSS_EDR_2002}. We use {\sc topcat} \citep{Topcat2005} to match the SDSS and GAMA galaxies, using the catalogue right ascension and declination values and requiring an on-sky separation of $<0.5\arcsec$. This results in a parent sample of 12,738 galaxies. 

The GAMA \magphys\ catalogue contains dust emission parameters even for galaxies with no mid- or far-IR photometric detections; these are predictions based on the \magphys\ model priors. We therefore restrict our analysis to only those galaxies with $\ge2$ photometric bands between 22--500\,\textmu{}m detected at $>3\sigma$ in the {\sc lambdar} catalogue. This reduces the sample to 4,368 galaxies. Finally, we restrict our sample to objects with \Ha detected at $>5\sigma$, leading to a final sample of 4,121 galaxies with SDSS spectroscopy, GAMA photometry,  H-ATLAS/WISE and \Ha emission line detections. 

\vspace*{2ex}

\subsubsection{Star-forming and retired samples}\label{sec:dataSFretired}

Of these 4,121 galaxies, we find that 1,638 galaxies have optical emission line ratios that indicate the gas is solely ionised by young stellar populations, using the criteria of \citet{Stasinska2006}, which is based on the commonly used \nii/\Ha and \oiii/\Hb line flux ratios. We find that 386 galaxies have \Ha equivalent widths $\WHa<3$\AA, the limit below which a galaxy is classified as `retired' \ie the emission lines are consistent with the ionisation source being hot low-mass evolved stars (HOLMES) present in the galaxy continuum spectrum \citep{CidFernandes2011}. We also find 305 galaxies are classified as AGN using the \citet{kewley2001} criteria. The ionising photons responsible for the \Ha\ emission in the remaining galaxies likely come from a mixture of HOLMES, ongoing star formation, shocks and low luminosity AGN. For the purposes of this paper it is essential to include galaxies with all ionisation sources, but we refrain from converting \LHa into SFR for our figures. Figure \ref{fig:sample} shows the samples in stellar mass vs. redshift, and on standard emission line diagnostic diagrams.

\subsubsection{Post-starburst sample}\label{sec:dataPSB}
There are many ways to identify post-starburst galaxies, although all methods are based around identifying galaxies with an excess of A/F stars compared to their O/B star population, indicating a recent and rapid shut-off in star formation. Traditionally a complete absence of nebular emission lines was used to provide evidence for a lack of O/B stars \citep[\eg][]{dressler1983,couch1987,zabludoff1996,Goto2003}. However, this method excludes the many interesting post-starburst galaxies with residual ongoing star formation and other ionisation sources. Therefore, most modern methods use an alternative method to determine a deficit of O/B stars relative to the A/F star population, such as information from the stellar continuum shape, the equivalent width of one of the emission lines, likely ionisation sources of the emission lines or longer wavelength photometry \citep{wild2007,Yan2009, alatalo2016, yesuf2017}. 

Due to our requirement for significant \Ha emission line flux, as well as mid- and far-IR photometric detections, we note that the post-starburst sample studied in this paper is biased to include only post-starburst galaxies with a significant interstellar medium, as well as sources to heat and ionise it, such as residual star-formation, shocks, HOLMES or AGN. However, as it is predominantly these galaxies which are argued to be dusty starburst interlopers in quenching samples \citep[\eg][]{PoggiantiWu2000,Baron2022}, they are the relevant objects for this analysis. 

\begin{table*}
    \centering
    \caption{Position, catalogue identifiers and basic properties of the post-starburst galaxies studied in this paper. }
    \label{tab:sample_psb}
{\renewcommand{\arraystretch}{1.3}% for the vertical padding
\begin{tabular}{cccccccccccr}
\hline
RA & Dec & Plate & MJD & Fiberid &  GAMA-DR4 & Catalogues & $z$ & \Mgal & \LHa & \Ldust & \WHa \\
$\mathrm{{}^{\circ}}$ & $\mathrm{{}^{\circ}}$ &  &  &  & CATAID &  &  & dex(M$_\odot$)  & dex(L$_\odot$) & dex(L$_\odot$) & \AA \\\hline

130.77146 & 1.14878 & 467 & 51901 & 397 & 300757 & 5 & 0.078 & 10.4 & 7.6 & 10.8 & 4.1 \\
133.40527 & 1.71763 & 469 & 51913 & 338 & 323854 & 2 & 0.058 & 10.7 & 8.3 & 10.4 & 7.7 \\
134.17632 & 1.03737 & 468 & 51912 & 557 & 372076 & 2 & 0.088 & 10.7 & 7.4 & 9.9 & 2.9 \\
135.19133 & -0.49369 & 470 & 51929 & 290 & 550327 & 2 & 0.041 & 9.5 & 7.6 & 10.1 & 24.0 \\
135.50417 & -0.56457 & 470 & 51929 & 251 & 550389 & 2 & 0.089 & 9.9 & 7.6 & 10.4 & 14.6 \\
135.88658 & 1.21010 & 470 & 51929 & 402 & 372446 & 2,4,5 & 0.058 & 9.8 & 6.1 & 9.5 & 0.6 \\
136.75280 & 0.06094 & 470 & 51929 & 115 & 210106 & 2,5 & 0.164 & 10.4 & 7.8 & 10.6 & 4.8 \\
137.33520 & 0.73873 & 470 & 51929 & 621 & 623161 & 2 & 0.054 & 10.2 & 7.6 & 9.8 & 12.5 \\
139.67269 & 2.02517 & 473 & 51929 & 464 & 383051 & 2 & 0.087 & 11.0 & 8.2 & 10.7 & 6.3 \\
140.82538 & 0.56146 & 474 & 52000 & 408 & 217011 & 2 & 0.088 & 10.4 & 7.9 & 9.9 & 3.6 \\
140.90214 & 2.13637 & 473 & 51929 & 593 & 383312 & 2 & 0.017 & 10.6 & 7.5 & 10.3 & 1.9 \\
140.92896 & 1.34917 & 473 & 51929 & 29 & 373374 & 2 & 0.025 & 11.0 & 6.7 & 10.0 & 1.2 \\
177.64780 & -2.63937 & 329 & 52056 & 117 & 123917 & 3 & 0.093 & 10.6 & 7.9 & 10.5 & 13.3 \\
177.65639 & 1.50016 & 515 & 52051 & 295 & 397294 & 1,2,4,5 & 0.078 & 10.2 & 6.5 & 9.9 & 1.0 \\
178.10642 & -1.26744 & 330 & 52370 & 361 & 143889 & 1,2,5 & 0.060 & 10.4 & 6.5 & 10.1 & 2.3 \\
179.98362 & -0.68304 & 285 & 51663 & 162 & 40161 & 4 & 0.080 & 10.3 & 7.6 & 10.1 & 10.6 \\
180.74100 & 0.05999 & 286 & 51999 & 395 & 70638 & 2,4 & 0.080 & 10.9 & 8.0 & 10.5 & 9.0 \\
181.11809 & -0.88879 & 286 & 51999 & 208 & 536202 & 2 & 0.094 & 10.0 & 8.4 & 10.4 & 43.5 \\
181.22794 & 0.31118 & 286 & 51999 & 477 & 610664 & 1 & 0.100 & 10.7 & 7.8 & 10.6 & 5.1 \\
181.43242 & 1.07895 & 286 & 51999 & 444 & 23386 & 4 & 0.086 & 10.7 & 7.8 & 10.5 & 9.2 \\
181.62258 & -0.55774 & 286 & 51999 & 150 & 560745 & 2 & 0.095 & 10.6 & 8.1 & 10.6 & 5.4 \\
182.43333 & -0.28872 & 286 & 51999 & 78 & 55486 & 2 & 0.160 & 10.2 & 8.6 & 11.0 & 54.5 \\
182.70929 & 0.47087 & 286 & 51999 & 634 & 85480 & 4 & 0.078 & 10.4 & 7.9 & 10.3 & 13.5 \\
183.03908 & -1.68483 & 332 & 52367 & 527 & 138549 & 2,5 & 0.103 & 10.2 & 6.4 & 9.8 & 0.7 \\
212.55941 & -0.57859 & 302 & 51688 & 110 & 567624 & 3 & 0.026 & 9.6 & 6.9 & 9.2 & 14.2 \\
213.29913 & -0.39941 & 304 & 51609 & 291 & 62531 & 3 & 0.126 & 10.8 & 8.1 & 10.7 & 14.6 \\
214.16447 & 1.30599 & 533 & 51994 & 55 & 227568 & 2 & 0.123 & 10.2 & 8.2 & 10.3 & 17.8 \\
214.19902 & -1.07059 & 303 & 51615 & 89 & 37000 & 2,3,5 & 0.125 & 11.1 & 8.7 & 10.6 & 2.6 \\
214.46337 & -0.13896 & 304 & 51609 & 216 & 592879 & 4 & 0.169 & 10.8 & 8.3 & 10.6 & 14.1 \\
216.59955 & 2.26955 & 535 & 51999 & 341 & 342615 & 2,5 & 0.105 & 10.7 & 8.2 & 10.6 & 12.6 \\
216.81288 & 0.93103 & 306 & 51637 & 321 & 106634 & 2 & 0.052 & 10.4 & 7.8 & 10.1 & 17.4 \\
221.69920 & -1.21974 & 920 & 52411 & 460 & 493601 & 2,5 & 0.083 & 10.4 & 8.3 & 10.4 & 5.1 \\
223.07965 & 0.13545 & 308 & 51662 & 638 & 79784 & 2 & 0.147 & 10.6 & 7.9 & 10.8 & 18.1 \\

%\dots & \dots & \dots & \dots & \dots & \dots & %\dots & \multicolumn{1}{c}{\dots} \\
\hline
\end{tabular}
\\
Catalogue references:
(1) \citet{Goto2007}
(2) \citet{wild2007};
(3) \citet{alatalo2016};\\
(4) \citet{Pattarakijwanich2016};
(5) C. Tremonti and \citet{chen2019}. 
}
\end{table*}

We match our combined SDSS+GAMA+H-ATLAS/WISE sample onto several SDSS DR7 post-starburst samples in the literature, identifying 33 post-starburst galaxies in total. Positions, catalogue identifiers and basic properties for these galaxies are given in Table \ref{tab:sample_psb}. The sample includes 3 galaxies from \citet{Goto2007}, 24 from \citet{wild2007}\footnote{We follow \citet{pawlik2018} and exclude dusty PCA-selected post-starburst candidates with \Ha/\Hb ratios $>6.6$. These are possible contaminants for this particular method, due to the impact of significant dust attenuation on the shape of the stellar continuum.}, 4 from \citet{alatalo2016}, 7 from \citet{Pattarakijwanich2016} and 9 applying the method described in \citet{chen2019} to SDSS DR7, referred to as the `Tremonti sample'. Some objects are identified using multiple methods, with the largest cross over being 8 objects contained in both the \citet{wild2007} and Tremonti samples. The selection criteria of each sample is summarised in Table \ref{tab:select_psb}. 

\begin{table*}[]
    \centering
    \caption{Comparison between the different selection methods of post-starburst galaxies included in this paper. In order to distinguish absorption from emission in this table, equivalent width values ($W$) follow the convention of positive for absorption and negative for emission. Note that in the remainder of the paper we use absolute equivalent width values for  
    $\WHa$ in emission. }
    \label{tab:select_psb}
    {\renewcommand{\arraystretch}{2}% for the vertical padding
    \begin{tabular}{p{0.22\linewidth} | p{0.09\linewidth} | p{0.3\linewidth} | p{0.3\linewidth}}
    Author & Survey & Balmer absorption lower limit & Ongoing SFR upper limit \\
    \hline
    \hline
    \citet{Goto2007} & SDSS DR5 &  
    $W_{\Hd} >5$\,\AA\ & $\WHa>-3$\,\AA\ and $W_{\oii}>-2.5$\,\AA\footnote{Lines measured with flux summing method \citep{Goto2003}.}\\
    \hline

    \citet{wild2007} & SDSS DR7 &
    \multicolumn{2}{p{0.6\linewidth}} { $\mathrm{PC2}>0.025$ and $\mathrm{PC1}<-1.5$ \hfill 
    
    (\ie excess Balmer absorption compared to 4000\,\AA\ break strength)\footnote{Additional cut on Balmer decrement \citep{pawlik2018}.}} \\
    \hline

    \citet{alatalo2016} & SDSS DR7 &  $W_{\Hd}>5$\,\AA\ & Line ratios consistent with shock ionisation and inconsistent with star-formation ionisation.\\   
    
    \hline
    \citet{Pattarakijwanich2016} & SDSS DR9\footnote{Sample restricted to DR7 only in this paper.} & Template fitting\footnote{Method from \citet{quintero2004}.} $\mathrm{A/(A+K)}>0.25$ and $W_{\Hd}>4$\,\AA\ & $W_{\oii}>-2.5$\,\AA \\
    \hline

    Tremonti sample; \citet{chen2019} & SDSS DR7 & 
    $\Hda > 3$\,\AA\footnote{Lick Index \citep{worthey1997}.} & $\WHa>-10$\,\AA\ and \hfill
    
    $\log_{10} |\WHa| < 0.23 \times \Hda -0.46$  \\
    \hline

    \end{tabular}
}
\end{table*}

The sample includes 7 post-starburst galaxies where optical emission lines indicate that the dominant ionising source is residual star-formation (below the demarcation line of \citealt{Stasinska2006}), 8 are consistent with being ionised by HOLMES ($\WHa<3$\,\AA), 6 ionised by AGN (above the demarcation line of \citealt{kewley2001} and $\WHa>3$\,\AA), and 10 lie between the \citet{Stasinska2006} and \citet{kewley2001} demarcation lines of the \nii/\Ha and \oiii/\Hb line ratios diagram \citep{BPT1981}. The post-starburst galaxies are shown with darker coloured symbols in Figure \ref{fig:sample}.

\newpage
\section{Methods}\label{sec:methods}
Here we summarise the relevant data analysis and interpretation methods used in this paper. 

\emph{Partial correlation coefficients}: Throughout the results section we use partial correlation coefficients to quantify the correlation between three variables. For variables ($x$, $y$ and $z$) we firstly compute the Spearman rank-order correlation coefficient between each: $\rho_{xy}$, $\rho_{xz}$ and $\rho_{yz}$. The partial correlation between $x$ and $z$, given $y$, is then: 
\begin{equation}
    \rho_{xz|y} = \frac{\rho_{xz} - \rho_{xy}\rho_{zy}}{\sqrt{(1 - \rho_{xy}^2)} \sqrt{(1 - \rho_{zy}^2)}}.
\end{equation}
The angle of the correlation with $z$, projected onto the $x$--$y$ plane ($\theta_p$), is given by:
\begin{equation}
    \tan{\theta_p} = \frac{\rho_{xz|y}}{\rho_{yz|x}}.
\end{equation}
An angle of $45^\circ$ indicates the $z$ quantity correlates equally with the $x$ and $y$ quantities; $0^\circ$ indicates the $z$ quantity correlates with $y$ alone; $90^\circ$ indicates the $z$ quantity correlates with $x$ alone;  $-45^\circ$ indicates the $z$ quantity correlates equally with the $x$ and with $y$, but an increase in $x$ corresponds to a decrease in $y$.

\emph{Expected luminosity ratio}: We compare our observed ratio between the luminosities of \Ha and total IR (TIR), to the ratio expected if both quantities measure SFR. Following the notation of \citet{KennicuttEvans2012},  $\log_{10}(\mathrm{SFR}) = \log_{10}(L_X) - \log_{10} (C_X)$, with SFR in M$_\odot$/yr and luminosities in erg/s. With $C=43.41$ for TIR (3--1100\,\textmu{}m) and $C=41.27$ for \Ha \citep{Hao2011, Murphy2011}, we obtain the expected relation for star-forming galaxies of:
\be
\log_{10}(L_{\rm TIR}) = \log_{10}(\LHa) + 2.14.
\ee
We note that these conversions have a number of assumptions underlying them, most notably of constant SFR for 10\,Myr for \LHa and 100\,Myr for \Ldust. We will discuss the implications of this approximation at length in Section \ref{sec:discussion}. 

\emph{\magphys\ fitting}: Although the \magphys\ catalogue was computed by the GAMA team and made available as part of their data release, it is important to understand the underlying dust emission model in order to interpret our results. \magphys\ is an energy balance code that assumes that all stellar light absorbed from the UV--optical wavelength range is re-emitted in the mid- and far-IR. It fits a two component dust emission model, with one contribution from dust in the stellar birth clouds, and the second from dust in the ambient ISM (see figure~2 in \citealt{daCunha2008_magphys}). Both components contribute to flux emitted by polycyclic aromatic hydrocarbons (PAHs) and hot mid-IR continuum. The birth clouds additionally contribute warm dust in thermal equilibrium, while the ambient ISM contributes cold thermal dust. Free parameters are the fraction of total IR luminosity contributed by dust in the ambient ISM, the temperature of the two thermal components, and relative fractions of light contributed by the different birth cloud components. 

We note that the total IR luminosity output by \magphys\ is calculated from the energy lost in the UV--optical wavelength regime, which is then distributed via the model between 3--1000\,\textmu{}m. Although \Ldust is necessarily model dependent due to the sparse sampling of the IR SED by observations, comparisons between methods indicate that this does not introduce significant errors \citep[\eg][]{Hunt2019}.

\emph{Modified black body fitting}: To ensure that our results are not dependent on the \magphys\ model, we independently fit the total IR luminosity with a modified black body, using the python MCMC package \emph{emcee} \citep{emcee2013}. We assume a fixed dust emissivity index $\beta=1.8$ and a dust mass absorption coefficient $\kappa_{850}=0.077 \mathrm{\,m^2 \,kg^{-1}}$ and include a mid-infrared power-law to account for hot dust, allowed to vary uniformly between $0.5<\alpha<5.5$ \citep[see also][]{Casey2012,Bourne2019}. We fit for a temperature range between 10 and 100\,K using a uniform prior. We obtain a very tight 1:1 correlation between our measurement of $L_{\rm TIR}$ and the value from \magphys, with a Spearman rank correlation coefficient of 0.99 when all 6 mid- to far-IR bands are detected at $>3\sigma$, decreasing slightly to 0.96 when only 3 bands are detected.

\emph{Aperture correction}: Finally, our \LHa\ values are measured within a 3\arcsec\ diameter aperture fibre, with atmospheric seeing further scattering the light beyond the fibre aperture. On the other hand the \magphys\ \Ldust\ values are measured from the global galaxy photometry. For our sample of 4,121 galaxies, the median amount of $z$-band light contained within the fibre is  0.27, with 16th and 84th percentiles of 0.18 and 0.37. This effect can be approximately corrected by dividing the spectral flux by the fraction of $z$-band light ($L_z$) contained within the fibre. This will be less accurate in the case of strong radial gradients in $\LHa/L_z$, however we checked that there were no redshift trends in any of our results which would suggest this to be a problem. We initially performed our analysis with \LHa\ values uncorrected for this aperture effect, and the results were qualitatively identical. The advantage of doing the aperture correction explicitly is that galaxies with differing surface brightness profiles are treated properly relative to one another. The fact that our final \Ldust/\LHa\ ratio for highly star-forming galaxies lines up so well with that expected in the literature, and there are no redshift trends in our results, gives confidence that this method is appropriate.

\section{Results}\label{sec:results}
\begin{figure*}
    \centering
    \includegraphics[width=0.49\linewidth]{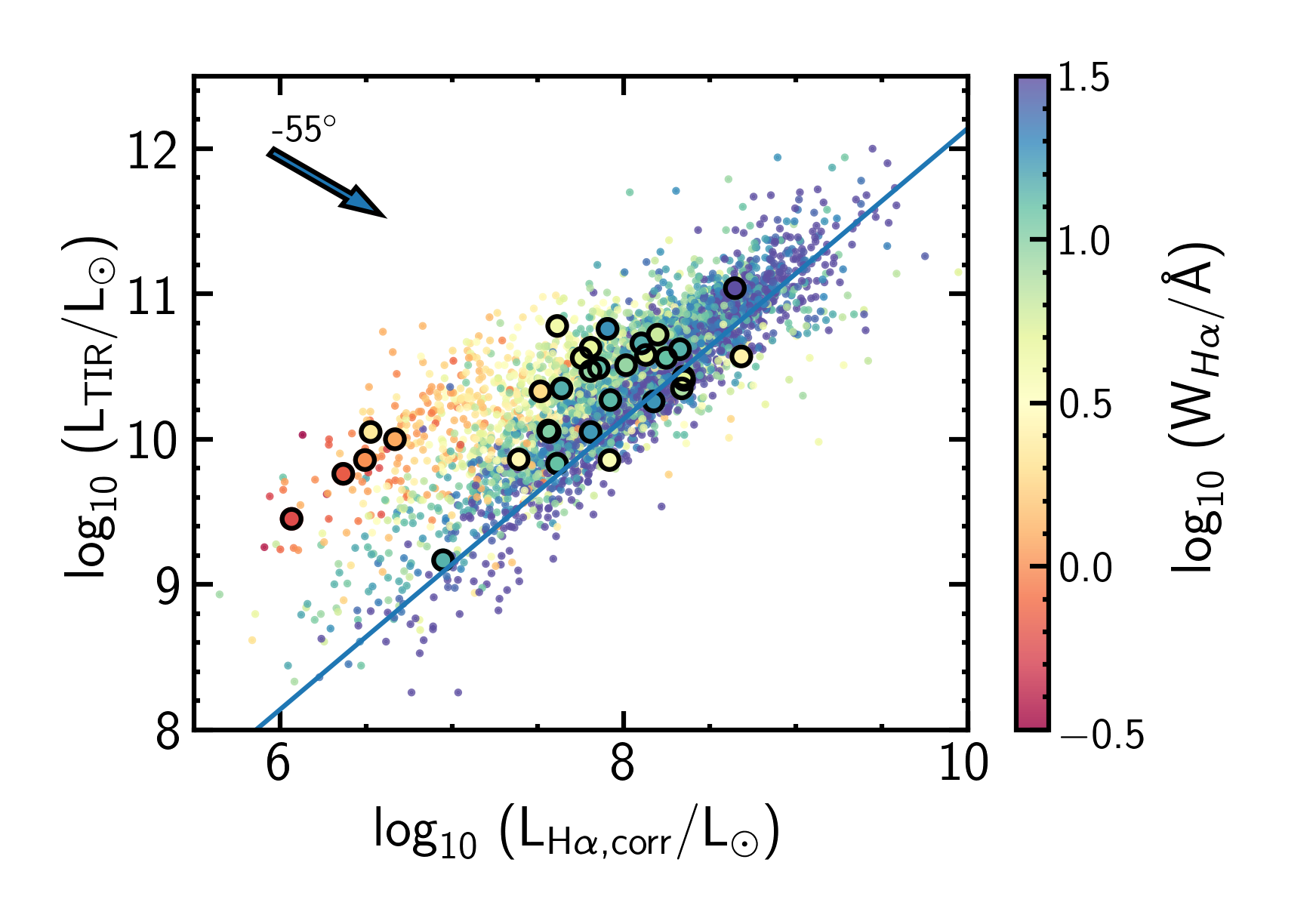}
    \includegraphics[width=0.49\linewidth]{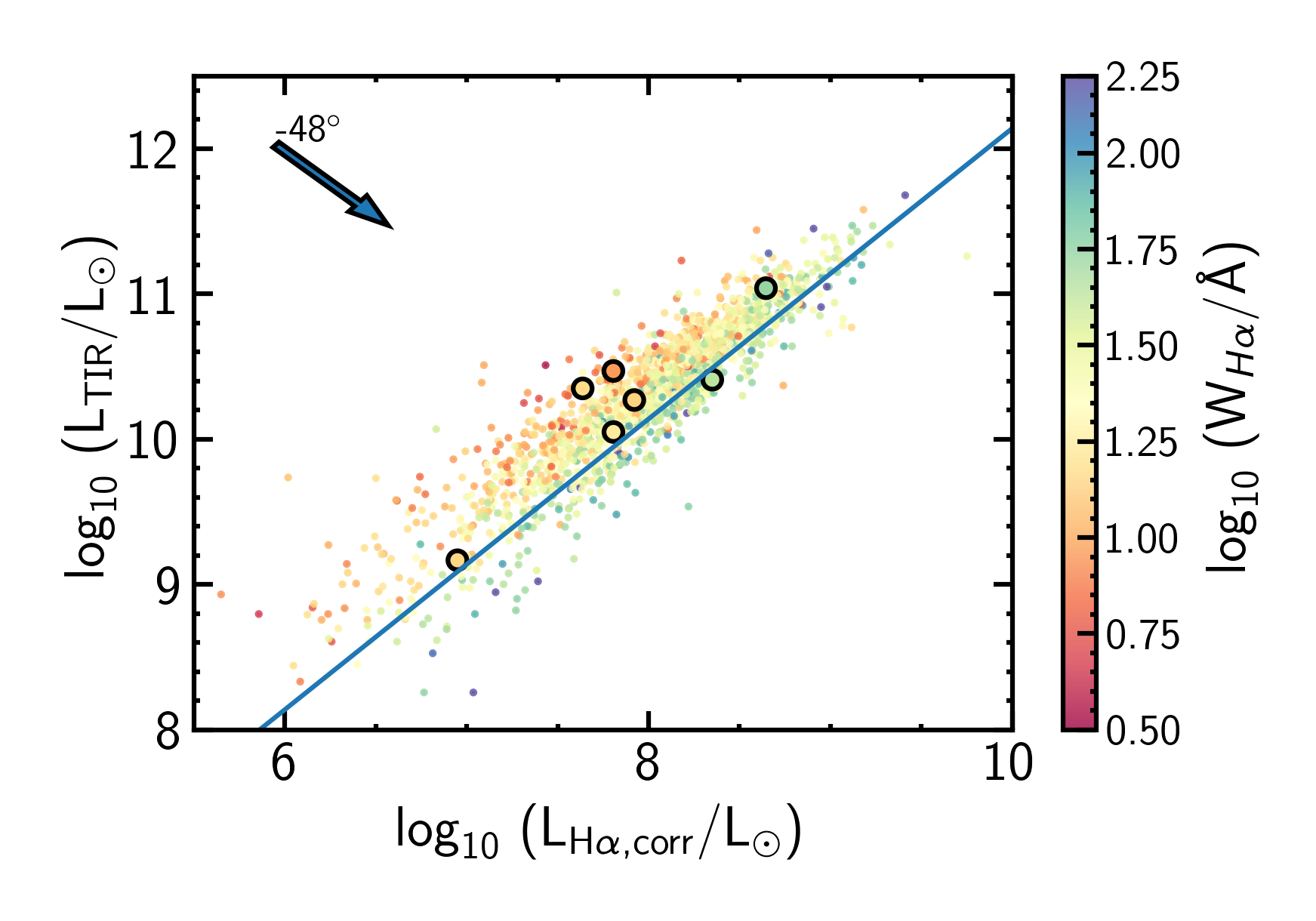}

    \caption{The total dust (mid-far infrared) luminosity versus \Ha luminosity for galaxies observed in both SDSS and GAMA/Herschel-ATLAS/WISE (small dots), where \LHa\ has been both dust and aperture corrected. Post-starburst galaxies are shown as large circles with the same colour scale. We only include galaxies with \Ha flux detected at $>5\sigma$ and fluxes in 2 or more 22--500\,\textmu{}m bands detected at $>3\sigma$. The arrow and angle indicate the direction of partial correlation with the $z$ (colour) axis. The line indicates equity in the case that both luminosities trace the galaxy SFR, assuming standard conversions from the literature. No smoothing has been applied to the data.
    {\it Left:} all galaxies, regardless of the origin of the \Ha flux. {\it Right:} only galaxies where the \Ha flux comes predominantly from star-formation, as determined from their emission line ratios (see text for details). 
    }
    \label{fig:LdustLHaEQW}
\end{figure*}

The left panel of Figure~\ref{fig:LdustLHaEQW} presents the total dust luminosity (\Ldust), as estimated by \magphys\ between 3--1000\,\textmu{}m, versus the dust and aperture corrected \Ha luminosity (\LHa) for all galaxies in our sample, colour coded by the equivalent width of \Ha (\WHa). We plot \WHa as this is a direct observable; for galaxies with emission lines dominated by star formation \WHa is strongly correlated with specific SFR ($\mathrm{sSFR}=\mathrm{SFR}/\Mgal$). If both \LHa and \Ldust were tracing SFR, galaxies should lie on the blue line. We can immediately see that this is only true for galaxies with \WHa$\gtrsim10$\AA. Retired galaxies with \WHa$\lesssim3$\AA\ (yellow to red points), lie significantly above or to the left the line, with higher \Ldust than expected for their \LHa.  This can not be explained by contamination of the \Ha line by additional ionisation sources such as AGN, HOLMES or shocks, as such contamination would scatter galaxies to the right of the blue line. The arrow indicates the direction of partial correlation ($\theta_p=-55^{\circ}$), confirming \WHa to be strongly positively correlated with \LHa, and negatively with \Ldust. 
%As we will see from the next plot, the stronger correlation with \LHa\ arises purely due to the presence of the retired galaxies. 

The right hand panel of Figure~\ref{fig:LdustLHaEQW} restricts the sample to the 1,638 galaxies with optical emission line ratios that indicate the gas is predominantly ionised by young stellar populations and therefore \LHa is a good proxy for SFR (Section \ref{sec:dataSFretired}).  It is noteworthy that there is still a trend with \WHa, with the conversion line equating both luminosities to SFR only holding for galaxies with $\WHa\gtrsim10$\,\AA\ \citep{Hao2011,Murphy2011,KennicuttEvans2012}. All other galaxies with lower \WHa lie above the line, with higher \Ldust than expected for their \LHa.  The partial correlation angle of $\theta_p=-48^{\circ}$ indicates that \WHa traces \Ldust\ just as much as it does \LHa and we can see that \emph{accounting for \WHa would account for most of the scatter in the relation}. The results shown in these figures are unchanged if we use a simple modified black body fit to the mid- and far-infrared data (see Section \ref{sec:methods}).

In both panels of Figure~\ref{fig:LdustLHaEQW} the post-starburst galaxies are indicated by larger circles. The identical colour scaling of the post-starburst galaxies relative to the background population of galaxies demonstrates that post-starburst galaxies are not unusual in their dust luminosities, given their \Ha equivalent widths. The majority of the post-starburst galaxies lie above the star-forming sequence in \Ldust versus \LHa. The small number with $\WHa\lesssim3$\,\AA\ have the highest excess \Ldust.

Our results show that all post-starburst galaxies have dust luminosities larger than predicted by \LHa, but consistent with star-forming galaxies or retired galaxies with the same \WHa. If these post-starburst galaxies are in fact `dust obscured starbursts', then the excess \Ldust common to \emph{all galaxies} with $\WHa<10$\,\AA\ should be interpreted as due to complete obscuration of the \Ha line. In Section \ref{sec:discussion} we will discuss why we do not think this is the case.

\begin{figure*}
    \centering
    \includegraphics[width=0.49\linewidth]{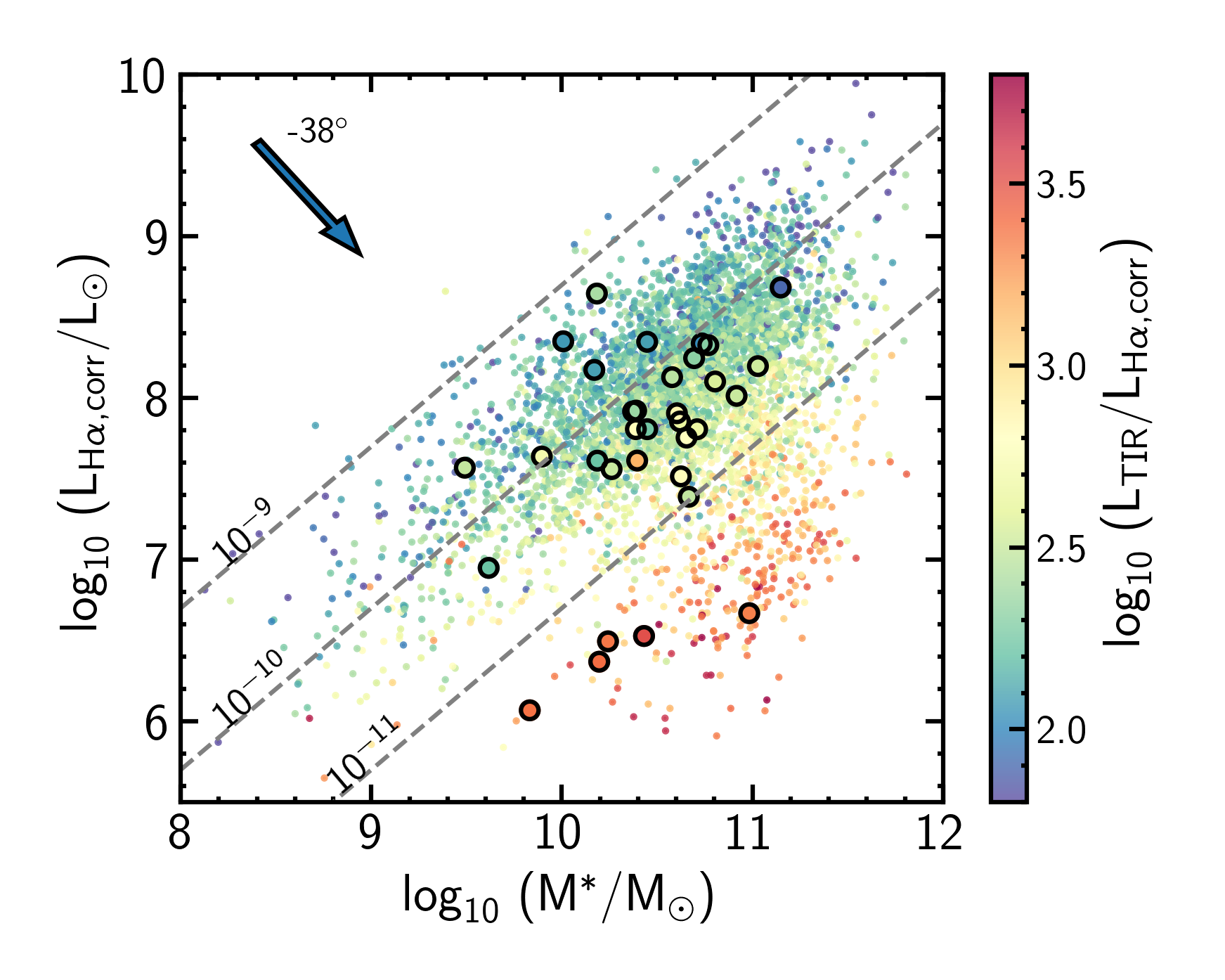}
    \includegraphics[width=0.49\linewidth]{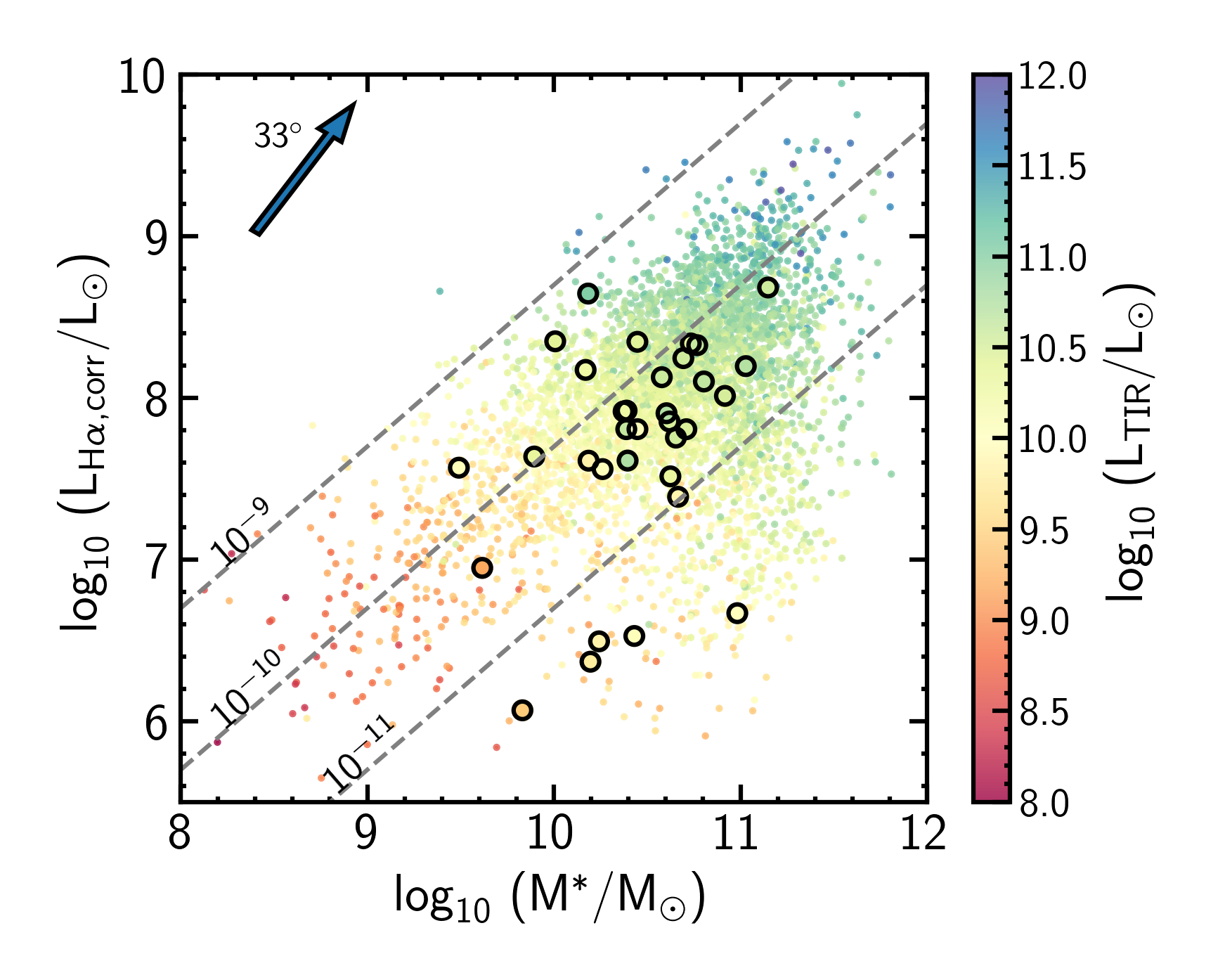}
    \includegraphics[width=0.49\linewidth]{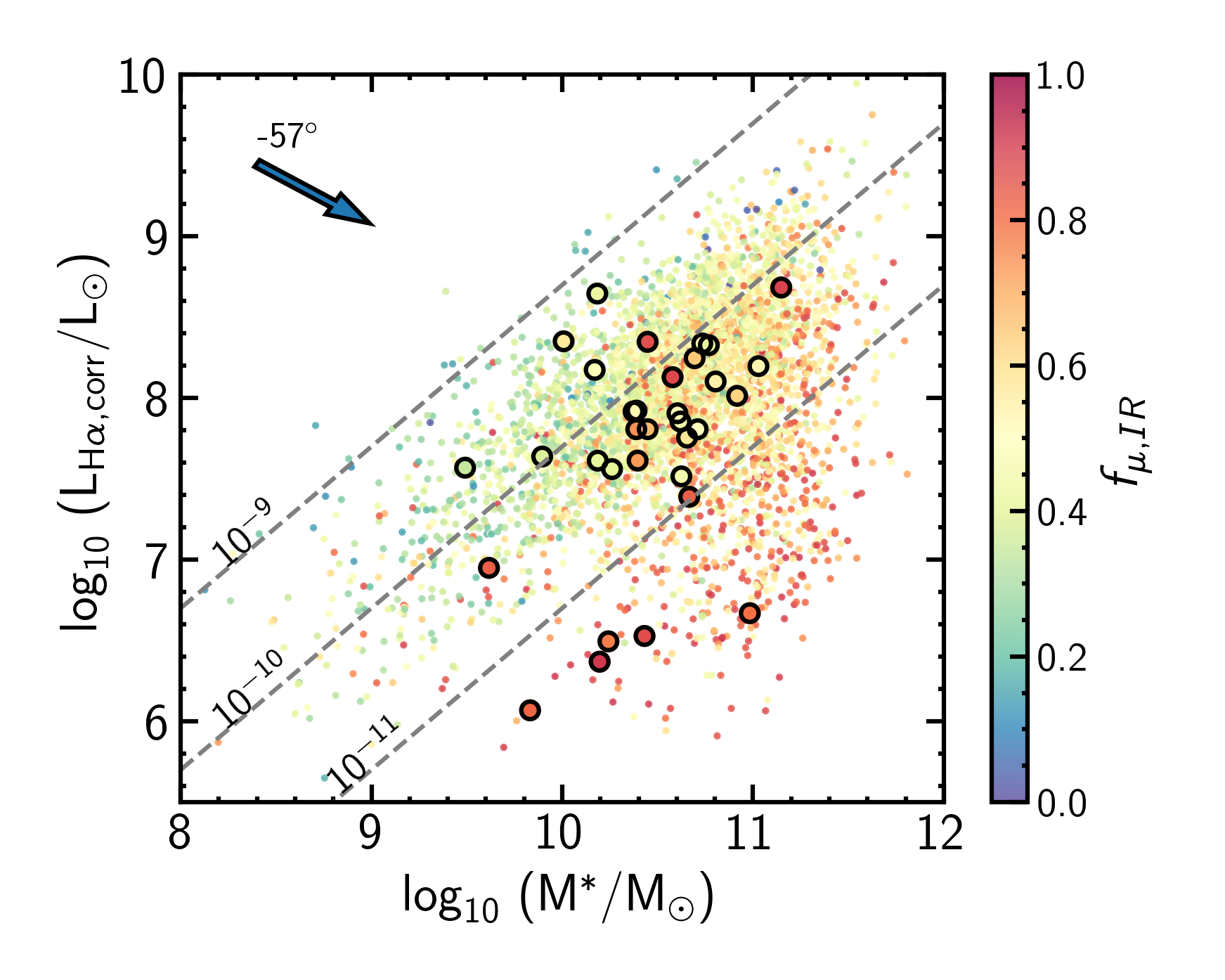}
    \includegraphics[width=0.49\linewidth]{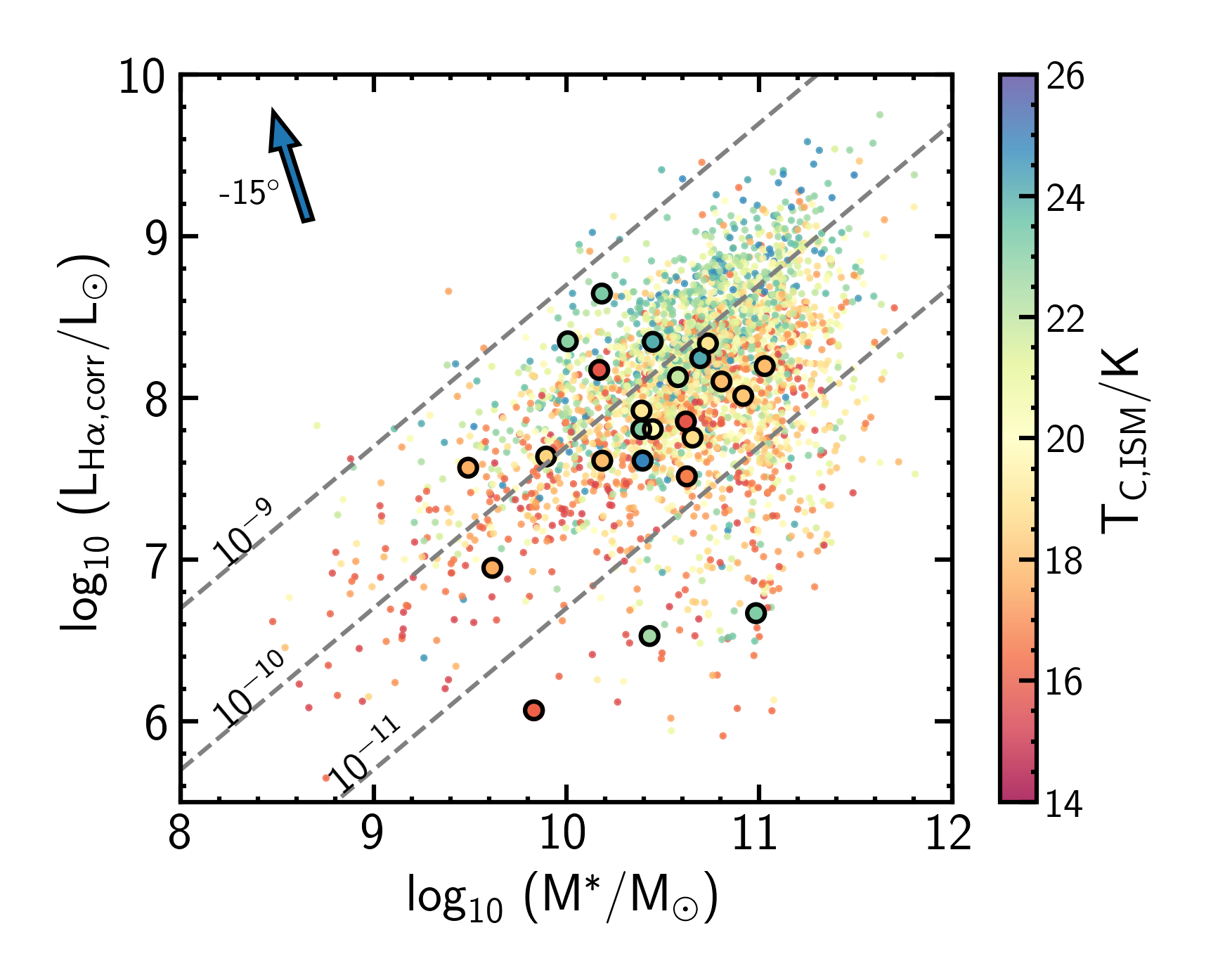}
    
    \caption{Total galaxy stellar mass versus \Ha luminosity for the same galaxy samples as in Figure \ref{fig:LdustLHaEQW}. \Ha luminosity has been corrected for dust attenuation and aperture, as described in the text. Dashed lines indicate constant sSFR as labelled, in units of yr$^{-1}$. Arrows and angles indicate the direction of partial correlation with the $z$-axis quantity. No smoothing has been applied to the data. \emph{Top left:} coloured by total infrared luminosity divided by dust and aperture corrected \Ha luminosity. The expected log(ratio) is 2.14 based on standard conversions from luminosity to SFR. \emph{Top right:} coloured by total infrared luminosity. \emph{Bottom left:} coloured by fraction of the total dust luminosity coming from the `diffuse' ISM, as fit by \magphys. \emph{Bottom right:} coloured by the temperature of the cool `diffuse' ISM dust component, as fit by \magphys. In this panel we require galaxies to have $>5\sigma$ detections in one additional dust band compared to our main analysis.
    }
    \label{fig:MstarHa}
\end{figure*}

To further investigate the cause of the varying \Ldust/\LHa ratio with \WHa, Figure \ref{fig:MstarHa} shows the same samples plotted as stellar mass \Mgal versus \LHa. We overplot lines of constant sSFR, although caution that these are only relevant where \LHa arises predominantly from star formation. The top left panel largely repeats the information presented in Figure \ref{fig:LdustLHaEQW}, but the inclusion of stellar mass makes the blue and red sequences stand out more clearly. Interestingly, the \Ldust/\LHa ratio changes smoothly across this figure, with no obvious discontinuity between the blue and red sequences, or for the post-starburst galaxies. The partial correlation angle of $\theta_p=-38^{\circ}$ is consistent with \Ldust/\LHa trending most closely with \LHa/\Mgal (or sSFR for galaxies where \LHa is dominated by star-formation). The smoothness of the transition suggests that \emph{whatever is responsible for the varying ratio has no knowledge of the origin of the emission, or whether the galaxy is quenched, recently quenched or still star-forming.} 

Red sequence galaxies clearly have the highest \Ldust/\LHa, with the majority having a ratio more than 1~dex greater than the expected ratio based on standard conversions (2.14, see Section \ref{sec:methods}), while 19 galaxies have a ratio $\ge 1.5$~dex greater.  If we accounted for contamination of the \Ha line from ionisation sources that are not related to star formation, such as AGN, shocks or HOLMES, this discrepancy would further increase. At the other extreme, galaxies with either very high \LHa/\Mgal (high sSFR for galaxies where \LHa is dominated by star-formation) or very high \LHa have a ratio that is lower than expected, with 32 having a ratio $\ge 0.5$~dex lower. Again we see that post-starburst galaxies, whether completely quenched or not, have the expected \Ldust/\LHa ratio compared to other galaxies with the same stellar mass and \LHa. 

The top right panel colours the galaxies by \Ldust. This visualisation adds no new data, but again demonstrates the continuity between the blue and red sequence. \Ldust depends almost equally on \LHa and \Mgal, \ie at fixed \LHa, \Ldust increases with \Mgal, again showing that both quantities cannot trace SFR. The smoothness of the colour scaling across the blue and red sequence again shows that \Ldust is entirely agnostic as to the origin of the emission or quenched status of the galaxy. 

The lower left panel colours the galaxies by the fraction of total dust luminosity that is fit by the ambient ISM dust component in \magphys ($f_{\mu, \mathrm{IR}}$), as opposed to the birth cloud dust surrounding the star-forming regions. We note that this result is dependent on the \magphys model, unlike previous results, but is useful to further interpret our results. Here we see a more distinct discontinuity between the blue and red sequences, with the majority of red sequence galaxies requiring $\lesssim10$\% additional contribution from a hotter birth cloud component, while the blue sequence galaxies have typically  $\gtrsim50$\% contribution from hotter birth clouds. 

The lower right panel colours the galaxies by the temperature of the ambient ISM dust component fit by \magphys. Constraining dust temperature accurately requires more dust bands than a simple dust luminosity, we therefore required galaxies to be detected at $5\sigma$ in 3 or more 22--500\,\textmu{}m bands for this analysis (2,877 galaxies)\footnote{For both the bottom left and bottom right panels, we verified that the results do not change if we require significant detections in more mid and far-IR bands.}. While there is a significant scatter, the predominant trend is for galaxies with higher \LHa to have higher temperature ISM dust ($\theta_p=-15^\circ$). There is a small residual trend with stellar mass, but not to the same extent as for \Ldust/\LHa. This shows that the changing ISM dust temperature is not the cause of the changing luminosity ratio. 
\magphys\ also fits for the temperature of the birth cloud thermal component (not shown). We find a similar result, with galaxies with higher \LHa having higher temperature birth cloud dust and any residual trend with stellar mass being small.  

We investigated various other correlations in the dataset, which are not shown for conciseness, but we describe a couple more here.  Dust attenuation, as measured by both the emission line Balmer decrement and by the \magphys\ multi-wavelength photometric fit, follows the same trend as \Ldust in blue sequence galaxies \ie\ higher mass, higher \LHa\ galaxies have greater attenuation. However, the trend does not extend to the red sequence, where there is lower attenuation overall ($\tau_{\rm V,Balmer}\lesssim0.5$). The relative luminosity fraction of the warm and cold thermal components fit by \magphys\ are correlated predominantly with stellar mass, with a small residual trend with \LHa. High mass galaxies and red sequence galaxies have significantly higher fraction of dust fit with a cold thermal component than lower mass, blue sequence galaxies.  This provides a consistent picture to that shown for $f_{\mu, \mathrm{IR}}$, indicating ambient
ISM dust heating depends on stellar mass via the amount of UV and optical photons available from lower mass stars, and therefore contributes most significantly to \Ldust in red sequence galaxies. 

\subsection{Post-starburst galaxies}
In the above figures it is tempting to look for indications that post-starburst galaxies might have different dust emission properties from the background galaxy population. Such a result might be expected given the commonly assumed origin for post-starburst galaxies as arising from  merger induced intense centralised starbursts \citep[although this origin may depend on sample selection, ][]{Ellison2024ArXiv}. We tested this by selecting a sample of 10 control galaxies for every post-starburst galaxy, matching with the closest Euclidean distance in \Mgal and \LHa. We then plotted the distributions of the parameters described in this section, and computed a Kolmogorov--Smirnov test statistic to search for statistically significant differences. Only the continuum optical depth as measured by \magphys\ showed a noteworthy, albeit statistically marginal, difference between the post-starburst and control galaxies ($D=0.25$, $p\text{-value} =0.04$), with the post-starburst galaxies having a slightly higher median value of $\tau_V=1.4$ versus $1.1$ for the mass and luminosity matched control sample. \emph{Our results conclusively show that there is no substantial difference in either the dust luminosity, or the dust properties, of post-starburst galaxies compared to galaxies with more normal star formation histories matched in \Mgal and \LHa.} 

\section{Discussion}\label{sec:discussion}
There are two scenarios by which \Ldust and \LHa should not track each other, which are interrelated, yet subtly different. Firstly, dust heating of older stars contributes more to \Ldust than to \LHa. The extent to which this matters depends on (a) the mass of older stars in the galaxy, and therefore on the star formation history of the galaxy in the distant past, and (b) the presence of ambient ISM that is available to be heated. Secondly, even in a star-forming galaxy, if the SFR rapidly decreases, \LHa will decrease rapidly while \Ldust will remain high for $\sim$100\,Myr. In this section we argue that dust heating by older stars, rather than a recent change in SFR, is the dominant effect. 

\subsection{A two-component ISM}
In the standard picture \citep[\eg][]{Helou1986, LonsdalePersson1987, DraineLi2007}, a galaxy ISM contains a temporally stable ambient reservoir of dust that is built up steadily via many generations of stars.  Alongside this ambient dust, there are temporally transient dense birth clouds surrounding new star-forming regions; the greater the number of star-forming regions, the greater the number of dense birth clouds. When these young star-forming regions dominate over the ambient dust, we find that \Ldust correlates strongly with \LHa and both provide a good estimate of SFR in galaxies. As the specific SFR of the galaxy declines, \Ldust becomes increasingly dominated by the ambient dust. This standard picture of dust in galaxies is the basis of the two component dust model of \citet{CharlotFall2000}, which is the model implemented in \magphys, as well as some other SED modelling codes \citep[\eg\ GRASIL,][]{Grasil3D2014}.

In fact, the sources of \LHa in galaxies can be thought of in much the same way as the sources of \Ldust: a background population of hot, low mass, evolved stars (HOLMES) provides the temporally stable equivalent of the ambient dust that builds up over time, while the temporally transient hot OB stars provide the majority source of Lyman continuum photons when present. Not entirely coincidentally of course, the timescale during which we expect OB stars to stop providing Lyman continuum photons is of the same magnitude as the timescale that we expect the dense birth clouds to disperse. Therefore, as sSFR decreases, the stellar population moves from OB star dominated to G+ star dominated, at the same rate that the dust clouds move from birth cloud dominated to diffuse ISM dominated. 

However, \Ldust and \LHa respond differently to the stellar population mix. Because \LHa emission caused by HOLMES is very weak, \LHa drops very sharply once the star-forming regions vanishes. On the other hand, \Ldust drops more slowly over a time period of 100\,Myr, and remains persistently higher than \LHa until the dust is destroyed, because dust can be heated by all stars emitting in the UV and optical wavelength regime (see figure 3 of \citealt{Hayward2014}).  Thus, the \Ldust/\LHa ratio increases and remains high after a decline in star formation, reflecting the higher fraction of dust luminosity coming from the ambient ISM compared to the fraction of \LHa coming from older stars.  We hypothesise that it is this effect that causes the smooth trend between \Ldust, \LHa and \Mgal continuing into the red sequence, without discontinuity, despite the apparently different ionising and heating mechanisms, and star formation histories of the galaxies. The changing  \Ldust/\LHa with \LHa/\Mgal then simply reflects the differing relative importance of OB stars to HOLMES for \LHa, and diffuse ISM to birth clouds for \Ldust emission.

As discussed in Section \ref{sec:methods}, when luminosities are converted into a SFR, a constant SFR over a fixed timescale is assumed (10\,Myr for \LHa and 100\,Myr for \Ldust). While this might be a reasonable approximation for a star-forming galaxy, it clearly is not for a young post-starburst galaxy where the SFR has dropped precipitously over the past hundred Myr. To untangle the relative importance of star formation history versus ambient dust heating to the excess \Ldust seen in retired, post-starburst and star-forming galaxies with $\WHa<10$\,\AA, we turn to toy models.

\begin{figure}
    \centering
    \includegraphics[width=1.0\columnwidth, trim=0 10 0 0]{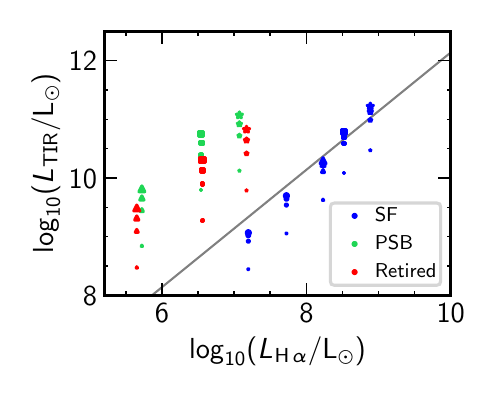}
    \caption{Total infrared versus \Ha luminosity for the star-forming (SF), post-starburst (PSB) and retired toy models in \Mgal bins. Triangles, squares and stars are used respectively for the middle, second-to-last and largest mass bins. The two lowest-mass bins for star-forming galaxies are shown as dots. Larger symbols indicate larger effective dust attenuation, with $\tau_V = 0.1, 0.5, 1.0$ and $2.0$. The grey line is the SFR conversion line as in Figure~\ref{fig:LdustLHaEQW}.} 
    \label{fig:toymodel-LHaLdust}
\end{figure}

\subsection{Toy SED models}
To illustrate how \LHa and \Ldust depend on a galaxy's stellar populations, we constructed a set of toy SED models based on average star formation histories for star-forming, post-starburst and retired galaxies in the \starlight--SDSS main galaxy sample catalogue. We note that the post-starburst galaxies were selected via the \citet{Goto2007} method, meaning they have completely ceased star-formation and thus match the subset of our post-starburst galaxies with $\WHa<3$\,\AA. The models were constructed in mass bins of 1\,dex using \citet{bruzual2003} models with a \citet{Chabrier2003} initial mass function. Further details are provided in Appendix \ref{app:SFH}, including a figure demonstrating the difference in star formation histories between the star-forming, retired and post-starburst samples (Figure \ref{fig:toymodel-SFH}).

We calculated \LHa from the extrapolated SED models in the UV by integrating the rate of ionizing photons, assuming the case B hydrogen recombination and an ionisation-bounded scenario \citep[as in][]{Stasinska2006, CidFernandes2011}. For quenched post-starburst and retired galaxies, since the measured light fraction from young stars is over-estimated due to limitations in the stellar population models, we only considered stellar populations older than $10^8$\,yr \citep[see][equation 2 and footnote 6]{Ocvirk2010, CidFernandes2011}.

From the intrinsic model spectrum $L_\lambda^\mathrm{int}$, we modelled a dust reddened spectrum assuming a single dust screen,
\be
L_\lambda^\mathrm{obs} = L_\lambda^\mathrm{int} e^{-\tau_V q_\lambda},
\ee
where $\tau_V$ is the effective dust attenuation in the $V$-band and $q_\lambda \equiv \tau_\lambda / \tau_V$ is the shape of the dust attenuation law, here taken to be the \citet{Cardelli.Clayton.Mathis.1989a} extinction curve with $R_V = 3.1$. We created models with four values of the effective dust attenuation, $\tau_V = 0.1, 0.5, 1.0$ and $2.0$. \Ldust was then calculated by integrating the difference between the intrinsic and dust reddened spectrum, and assuming all light attenuated by dust is re-radiated in the mid- and far-IR.

Figure~\ref{fig:toymodel-LHaLdust} shows the results of our toy models in the \LHa--\Ldust plane, which closely match those seen in Figure~\ref{fig:LdustLHaEQW}. Larger symbol sizes indicate greater $\tau_V$ values, which affect only \Ldust, since the intrinsic and dust-corrected \LHa are taken to be the same. Triangles, squares and stars are used respectively for the middle, second-to-last and largest mass bins. \emph{Crucially, only models for star-forming galaxies fall onto the SFR conversion line, with retired and quenched post-starbursts lying up to 1.5~dex above it, supporting our conclusion that \Ldust is only suitable as an SFR indicator in galaxies that are actively star forming.} 

When going from models for star-forming galaxies to post-starburst/retired galaxies, within the same mass and $\tau_V$ bin (fixed symbol style and size, changing colour) there is a slight decrease in \Ldust as old stars still heat the diffuse ISM, and a huge drop in \LHa. This highlights the fact that HOLMES are a much less powerful ionising source than OB stars: the rate of ionising photons decreases by five orders of magnitude from a 10\,Myr to a 1--10\,Gyr stellar population (see figure 1, panel b of \citealt{CidFernandes2011}). 

The slightly higher \Ldust of the quenched post-starburst galaxies compared to retired galaxies is due to the presence of a large number of relatively hot A and F stars to heat the dust, rather than the recent precipitous drop in SFR (which from Figure~\ref{fig:toymodel-SFH} can be seen to be insignificant for well over the 100\,Myr averaging time assumed for \Ldust in star-forming galaxies). The comparatively small difference between post-starburst and retired galaxies, compared to the difference between post-starburst and star-forming galaxies, shows that \emph{the recent star formation history is a much less important factor than the specific star formation rate in determining the \Ldust/\LHa ratio}.

Our toy models do not include post-starburst galaxies that are in the process of quenching, which are seen in Figure~\ref{fig:LdustLHaEQW} to lie just above the star-forming sequence. Here we expect a combination of a rapidly declining recent SFR and a reduction in birth cloud dust to be contributing to the enhanced \Ldust compared to \LHa \citep[see also][]{Hayward2014}. We also do not include any timescale for dust destruction, so our toy models are only pertinent to galaxies which have some ambient ISM available to be heated.

\subsection{Dust-star geometry, interstellar radiation field, dust destruction and cloud density}
Interestingly, the many complications that could arise in terms of dust temperature, dust-star geometry, dust destruction and dust cloud density and even ionisation source do not appear to be important to our results, at least to first order, even for post-starburst galaxies which should provide a test of the most extreme quenching conditions. At far-IR wavelengths, the temperature of the dust is thought to be determined mostly by the dust-star geometry, rather than the shape of the interstellar radiation field \citep[ISRF, \eg][]{Bernard2010}. Thus, for any given stellar population mix, the stars which emit the most UV--optical photons will contribute most to the total \Ldust, \ie\ OB stars in star-forming galaxies, AF stars in post-starburst galaxies, or G and later in retired galaxies, but the differing temperature of the stars is not relevant. The \magphys SED modelling results shown in Figure~\ref{fig:MstarHa} suggest that dust temperature correlates most strongly with \LHa, with a small residual mass trend. \magphys modelling also suggests that galaxies with the highest \LHa in our sample have higher ambient ISM and dense birth cloud dust temperatures. This is expected to be due to the closer proximity of the dust to the star-forming regions, rather than the changing ISRF \citep[\eg][]{Nersesian2019}. These particular results will likely be sensitive to the assumed model, and deserve further exploration. 

The lack of any appreciable difference in the temperature of post-starburst galaxy ISM dust compared to normal galaxies is interesting (bottom right panel of Figure~\ref{fig:MstarHa}), as we might expect the A/F stars to be co-located with the majority of the ISM dust following the end of the starburst, and therefore able to enhance the heating in a similar way to the brightest starbursts.  \citet{rowlands2015} found a decrease in post-starburst dust temperature with time since burst. However, they were studying generally younger post-starburst galaxies than in this sample, and temperature was estimated from colour index rather than the \magphys fit as used here. It seems likely that the dust heating observed in the youngest post-starbursts in their sample was due to a changing balance of hot birth cloud dust to cooler ISM dust following the star-formation episode, rather than a more general heating of the ambient ISM \ie closer to the $f_{\mu, \mathrm{IR}}$ parameter in the bottom left panel of Figure~\ref{fig:MstarHa}. Over time and with no subsequent star formation we expect the dust to be destroyed, and here post-starburst galaxies provide a valuable timescale that can be used to constrain the mechanisms responsible for dust destruction. \citet{zhihui2019} used the unique clock provided by post-starburst galaxies to measure a timescale for dust destruction of $\sim$250\,Myr. By 1\,Gyr post-burst, the majority of post-starburst galaxies have no detectable dust. 

\subsection{SFR estimators and full spectral fitting}
The debate about the importance, or otherwise, of dust heating by lower mass stars has a long history, with both sides likely `right' in their context. In early works, the inability to correct \LHa\ for stellar absorption and dust attenuation led to the conclusion that IR was a better total SFR indicator for galaxies, regardless of what was heating it \citep[\eg][]{LonsdaleHelou1985}. However, this argument became obsolete once better optical data became available \citep[\eg][]{RosaGonzalez2002, kewley2002b} and the two were shown to be exceedingly tightly correlated in local star-forming galaxies.   

Another argument is that it does not matter where the photons come from, they still represent star formation, and should be counted even if the star formation happened long ago. This is because combining \Ldust with a UV luminosity to account for the unextincted photons accounts for most photons generated by stars that are still alive at the time of observation, regardless of their mass or when they were formed. However, this argument is only plausibly useful when integrating in both time and space, not when trying to estimate the ongoing SFR of a particular galaxy, or set of galaxies, as it does not tell us the number of stars recently formed. 

A more important, yet subtle, discussion is around the timescales for different SFR estimators. To obtain the current SFR of a galaxy, the basic underlying principle is to count the number of stars just formed. This is, however, not easy for spatially unresolved data -- in reality we have to count the brightest, and therefore hottest, stars formed, then assume an initial mass function to give us the SFR. Unfortunately, even this is hard to do, so we resort to proxy estimators that are necessarily time averaged. 

For \LHa the time-averaging is the main sequence lifetime of O and B stars, about 10\,Myr. It is this short timescale that makes it attractive as a robust SFR indicator. For FIR, the time averaging is around 100\,Myr for star-forming galaxies, which accounts for lower mass stars also heating the dust \citep{kennicutt1998}.  \emph{All} conversions between a proxy luminosity and SFR require the assumption of a constant SFR over the averaging timeframe. For some galaxies, such as those which are currently rapidly declining in SFR, the longer estimators may be less suitable. Our results, supplemented by our toy models, highlight the subtle difference between star formation history versus contamination by ambient ISM dust in enhancing \Ldust/\LHa. The post-starburst galaxies in our sample that are currently decreasing in SFR, but still have appreciable SFR, have a slightly enhanced \Ldust for their \LHa, and they lie close to ordinary star-forming galaxies. This could be due to their rapidly declining star formation history, as well as their reducing contribution of birth cloud dust to \Ldust. On the other hand, the post-starburst galaxies that have completely quenched their SFR have a very substantially enhanced \Ldust for their \LHa, causing them to lie close to the retired galaxies.  When the observed luminosity is dominated by photons from significantly lower mass stellar populations, we are no longer measuring the number of recently formed stars, but rather the integrated mass formed during the entire lifetime of the galaxy. This leads to a strong correlation between \Ldust and \Mgal, rather than \Ldust and SFR. This suggests that we should see different partial correlation angles between \Ldust, \Mgal and \LHa for star-forming, retired and post-starburst galaxies. Unfortunately the small numbers of retired and post-starburst galaxies in our samples made it impossible to tease these relations apart, and a larger set of retired and quenched post-starburst galaxies would be required.

In principle, obtaining an SFR from full multi-wavelength SED fitting, rather than simple proxy estimators, should solve these problems \citep{Hunt2019}. However, this requires the underlying modelling assumptions to represent reality, both in terms of the star formation histories and the dust emission models. \citet{Hunt2019} shows that different SED fitting models and proxy indicators disagree the most when sSFR is low, suggesting that not all SED fitting models include the correct assumptions (or model `priors') in this regime.

\subsection{Comparison to previous work}
It has long been known that dust heating by old stars will impact low sSFR galaxies more than their high sSFR counterparts, with many cautionary notes in the literature \citep[see \eg][]{kennicutt1998}. In this section we compare our results to previous studies where mid- and far-IR luminosities have been noticed to apparently over-estimate SFR in galaxies.

Although papers purposefully studying SFR calibrators commonly investigate galaxy types, few provide measurements of the \WHa\ or sSFR of the galaxies being used for calibration. One exception is \citet{Hao2011} where the majority of galaxies have $\WHa>10$\,\AA, and their figure 7 shows an overestimation of the SFR using an FIR-based calibrator by $0.3$~dex for the small number of galaxies with $\WHa>5$\,\AA. This is entirely consistent with our results. 

\citet{Salim2016} presented the GALEX-SDSS-WISE Legacy catalogue, noting that galaxies with $\log_{10}\mathrm{(sSFR/yr}^{-1})<-11$ have mid-IR SFRs biased high by up to 2 dex, attributing this to the dust being heated primarily by old stars. Similarly, \citet{Rosario2016} looked at \LHa versus \Ldust in an early Herschel-SDSS sample similar to that studied here, and while they did not link their results to star formation histories, their plots show the same excess of \Ldust in low sSFR galaxies. \citet{Kouroumpatzakis2023} used the SED modelling code CIGALE \citep{boquien2019} to show that mid-IR-based SFR estimators overpredict SFR compared to CIGALE for galaxies with \WHa$<10$\AA. It is reassuring that entirely different analyses have lead to the same quantitative results.  

Our results are also consistent with those of \citet{Nersesian2019} who used CIGALE to note that galaxies with $\log_{10}\mathrm{(sSFR/yr}^{-1})>-10.5$ had dust mainly heated by radiation emitted by the young population, whereas lower sSFR galaxies had dust heated by the older population. This corresponds to $\WHa>10$\,\AA, matching our results. Recently \citet{Parente2024} have used a semi-analytic galaxy evolution model coupled to the GRASIL radiative transfer model to demonstrate the dust content of galaxies transitioning through the green-valley remains high following star formation quenching, leading to relatively large 250\,\textmu{}m  fluxes in agreement with GAMA observations. 

And finally, our results are consistent with the spatially resolved PHANGS--MUSE SFR calibration comparisons of \citet{Belfiore2023}, who noted a systematic increase in IR-based SFR estimators compared to dust attenuation corrected \LHa at low  specific star formation rates of    $\log_{10}\mathrm{(sSFR/yr}^{-1})<-10$.

\emph{All these results show that IR-based SFR indicators should only be used where $\WHa>10$\,\AA\ or $\log_{10} \mathrm{(sSFR/yr}^{-1})>-10.5$.}

\section{Summary}\label{sec:summary}
We have compared the \Ha luminosity, corrected for dust and aperture effects, to the total IR luminosity of 4,121 galaxies observed in both spectroscopic SDSS DR7 and GAMA DR4 catalogues, with detections in the mid- and far-IR observations from WISE and Herschel--ATLAS. We find that:
\begin{itemize}
    \item \Ldust is strongly correlated with \LHa for galaxies with \WHa$\gtrsim10$\AA\ or $\log_{10} \mathrm{(sSFR/yr^{-1})} >-10.5$, with a ratio that is consistent with locally determined calibrations. In this regime, both \LHa\ and \Ldust\ can be used to estimate integrated SFRs of galaxies. 
    
    \item However, for galaxies with $\WHa \lesssim 10$~\AA\ or $\log_{10} \mathrm{(sSFR/yr^{-1})} <-10.5$, \Ldust/\LHa increases by up to 1.5~dex. In this regime, \Ldust should not be used to measure the integrated SFRs of galaxies. We remind readers that \LHa can only be used to estimate the SFR of galaxies when $\WHa \gtrsim 3$~\AA\ \citep{CidFernandes2011}.

    \item The smoothness in the relationships between \Ldust, \LHa and galaxy stellar mass for galaxies, regardless of their dominant heating or ionisation source, is consistent with the `standard' 2-component picture for dust emission in galaxies. A background of old, low mass stars both heat the dust and provide a low level background of ionising photons in all galaxies, regardless of sSFR. As sSFR increases, the number of star-forming regions increases, and these then dominate the heating and ionisation budget.
    
    \item The reason for the increasing  \Ldust/\LHa ratio with decreasing sSFR is simply due to the fact that dust is heated by UV--optical photons, while \LHa relies on Lyman continuum photons alone. Thus, \LHa decays rapidly once star formation ceases with old stars providing only a very small number of far-UV photons, while dust continues to emit, presumably until the grains are destroyed. 

    \item We use toy models to show that the excess \Ldust is predominantly due to the increasing dominance of old, low mass stars heating the dust, and not due to a recent decline in star formation rate, combined with the assumption of constant star formation rates in the calibration conversions of luminosity into SFR. However, the rare population of galaxies with ongoing, yet rapidly declining, star formation rates (young post-starburst galaxies) may have slightly enhanced \Ldust simply due to the longer averaging time of \Ldust than \LHa when converted into an SFR.
    
    \item Recently and rapidly quenched local post-starburst galaxies do not show any evidence of hosting dust-obscured starbursts. Their high \Ldust/\LHa ratios are entirely consistent with being caused by the above processes.  
    
\end{itemize}   

An improved understanding of heating, ionisation and dust destruction mechanisms in retired and post-starburst galaxies could be gained from spatially resolved observations, both in the mid- and far-IR and with optical integral field spectroscopy. Sensitive radio observations would also provide additional assurance that significant amounts of dust obscured star formation was not being missed from the optical census of the Universe. However, our results indicate that the high fraction of rapidly quenched galaxies detected in optical surveys is not a result of high levels of contamination from dust-obscured starbursts, and therefore understanding rapid quenching pathways to quiescence remains of strong importance to understanding galaxy evolution.

\section*{Data Availability}
The SDSS DR7 data is available from \url{http://classic.sdss.org/dr7/}. The GAMA DR4 catalogues are available from \url{https://www.gama-survey.org}. \starlight catalogues are available on request to NVA or VW. 

% Acknowledgements
\section*{Acknowledgements}

We would like to thank Rob Kennicutt and Desika Narayanan for their help and advice in interpreting the results, and Kenneth Duncan, Ho-Seong Hwang and Leah Morabito for sharing data and information on datasets during preliminary investigations carried out prior to this paper being written. We would also like to thank Ian Smail and the anonymous OJAp referees who provided thoughtful and helpful feedback that improved this manuscript.

VW acknowledges Science and Technologies Facilities Council (STFC) grants ST/V000861/1 and ST/Y00275X/1. NVA and VW acknowledge the Royal Society and the Newton Fund via the award of a Royal Society--Newton Advanced Fellowship (grant NAF\textbackslash{}R1\textbackslash{}180403). NVA acknowledges support from Conselho Nacional de Desenvolvimento Cient\'{i}fico e Tecnol\'{o}gico (CNPq). KR acknowledges support from NASA grant 80NSSC23K0495.

GAMA is a joint European-Australasian project based around a spectroscopic campaign using the Anglo-Australian Telescope. The GAMA input catalogue is based on data taken from the Sloan Digital Sky Survey and the UKIRT Infrared Deep Sky Survey. Complementary imaging of the GAMA regions is being obtained by a number of independent survey programmes including GALEX MIS, VST KiDS, VISTA VIKING, WISE, Herschel-ATLAS, GMRT and ASKAP providing UV to radio coverage. GAMA is funded by the STFC (UK), the ARC (Australia), the AAO, and the participating institutions. The GAMA website is \url{https://www.gama-survey.org/}.

Funding for the Sloan Digital Sky Survey IV has been provided by the Alfred P. Sloan Foundation, the U.S. Department of Energy Office of Science, and the Participating Institutions. SDSS-IV acknowledges support and resources from the Center for High-Performance Computing at the University of Utah. The SDSS web site is \url{www.sdss.org}. 

SDSS-IV is managed by the Astrophysical Research Consortium for the Participating Institutions of the SDSS Collaboration including the Brazilian Participation Group, the Carnegie Institution for Science, Carnegie Mellon University, the Chilean Participation Group, the French Participation Group, Harvard-Smithsonian Center for Astrophysics, Instituto de Astrof\'isica de Canarias, The Johns Hopkins University, Kavli Institute for the Physics and Mathematics of the Universe (IPMU) / University of Tokyo, Lawrence Berkeley National Laboratory, Leibniz Institut f\"ur Astrophysik Potsdam (AIP),  Max-Planck-Institut f\"ur Astronomie (MPIA Heidelberg), Max-Planck-Institut f\"ur Astrophysik (MPA Garching), Max-Planck-Institut f\"ur Extraterrestrische Physik (MPE), National Astronomical Observatories of China, New Mexico State University, New York University, University of Notre Dame, Observat\'ario Nacional / MCTI, The Ohio State University, Pennsylvania State University, Shanghai Astronomical Observatory, United Kingdom Participation Group, Universidad Nacional Aut\'onoma de M\'exico, University of Arizona, University of Colorado Boulder, University of Oxford, University of Portsmouth, University of Utah, University of Virginia, University of Washington, University of Wisconsin, Vanderbilt University, and Yale University.

\bibliographystyle{mnras} % MNRAS
\bibliography{biblist}

\appendix{}

\section{Star formation histories for toy SED models}
\label{app:SFH}
\begin{figure}
    \centering
    \includegraphics[width=0.8\columnwidth, trim=20 0 0 0]{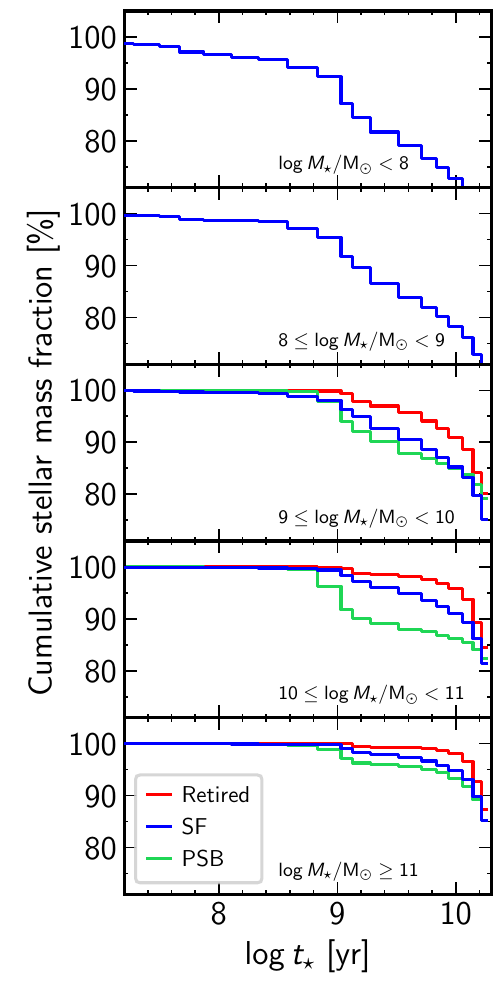}
    \caption{Average star formation histories for the retired, post-starburst (PSB) and star-forming (SF) subsamples in stellar mass bins, used in the toy models in Section \ref{sec:discussion}. 
    The cumulative star formation rate is shown as function of the stellar population age $t_\star$.
    %The amount of mass formed as a function of stellar population age $t_\star$ is represented by the initial mass-fraction population vector $\mu_\mathrm{ini}$.  
    } 
    \label{fig:toymodel-SFH}
\end{figure}

To create our toy SED models we selected galaxies from the \starlight--SDSS catalogue which belong to the SDSS main galaxy sample, with a signal-to-noise ratio $S/N \ge 10$ at 4020\,\AA, $\Mgal \ge 10^7$\,{M$_\odot$} and non-negative Petrosian half-light radius, which results in a parent sample of 616,097 objects. The star-forming sample comprises 133,769 galaxies below the \citet{Stasinska2006} line and with $\WHa \ge 3 \,\mathrm{\AA}$. Our subsample of 314,880 retired galaxies is defined as those with $\WHa < 3 \,\mathrm{\AA}$, while the 580 post-starburst were selected using the \citet{Goto2007} criteria.

Figure~\ref{fig:toymodel-SFH} shows the average \starlight fitted star formation histories (SFH) of our star-forming, post-starburst and retired galaxy subsamples split into five stellar mass bins. These are displayed in cumulative form, in order to best highlight the differences between the different subsamples. These average star formation histories are used to calculate \Ldust and \LHa in our toy models, as described in Section \ref{sec:discussion}. We excluded the two lowest mass bins for the retired and post-starburst samples, due to the low number of galaxies or very low predicted luminosities. 

We can see that retired galaxies have been quiescent for at least 1\,Gyr, at all stellar masses. Post-starburst galaxies experienced a significant burst of star formation between 500 and 1.5\,Gyr ago and have undergone minimal star formation since. Low mass star-forming galaxies show substantial star formation within the past 1\,Gyr, while higher mass galaxies formed an increasingly insignificant proportion of their mass recently. These star formation histories demonstrate that the star formation rate of all galaxies with stellar masses larger than $10^9$M$_\odot$, as in our samples (Figure \ref{fig:sample}), has been very stable for the last 100\,Myr. This is important for our discussion of the cause of the excess \Ldust in the retired and post-starburst galaxies in Section \ref{sec:discussion}. 

Our conversion from star formation history into \Ldust / \LHa depends on many properties of the stellar population models and dust attenuation curve, and the magnitude of the effect in Figure~\ref{fig:toymodel-LHaLdust} should only be taken as indicative. We investigated the impact of changing the assumed attenuation curve to a shallower \citet{CharlotFall2000} dust law, with $q_\lambda =  (\lambda/{5500\,\mathrm{\AA}})^{-0.7}$. This results in a modest increase in the predicted $\log \Ldust$ of 0.16~dex for the star-forming and of 0.08~dex for the post-starburst and retired galaxies. More recent stellar population models than the BC03 models used here yield slightly different SFHs when fit to optical spectra, \eg a newer version of the Bruzual \& Charlot models which include the new MILES stellar spectral library result in smoother SFHs \citep[\eg][]{pawlik2018}. Improvements in UV modelling post-\citet{bruzual2003} may slightly shift our results, but the key challenge lies in accurately characterizing young stellar populations. From a technical perspective, incorporating UV constraints into optical spectral synthesis can help to better constrain these populations \citep[\eg][]{LopezFernandez.etal.2018a, Werle.etal.2019a}. However, even the latest stellar libraries face limitations, particularly in their coverage of the crucially important hot young stars. Excluding stellar populations younger than $10^8$ years from our calculations is not ideal, as these do contribute some light to the galaxy SED, if not much mass. However, excluding these ages is a common approach taken to account for these limitations, with some stellar population models only starting at 30--60\,Myr \citep[\eg][]{Vazdekis2010}.  

\vspace{1cm}
\vfill

\end{document}